\documentclass[twocolumn,english,aps,prb,floatfix,rs,showpacs,amsfonts,amssymb,superscriptaddress]{revtex4-1}
\usepackage[T1]{fontenc}
\usepackage[latin1]{inputenc}
\usepackage{amsmath}
\usepackage{amssymb}

\makeatletter

\providecommand{\LyX}{L\kern-.1667em\lower.25em\hbox{Y}\kern-.125emX\@}

\usepackage{graphicx}
\usepackage{amscd}
\usepackage{babel}

\usepackage[compat=1.1.0]{tikz-feynman} 
\usepackage{caption}
\usepackage{subcaption}

\usepackage{bm}
\usepackage{ifthen}             
\newboolean{BookVersion} \setboolean{BookVersion}{False}    

\def\Rb87{^{87}\rm{Rb}}                     
\def\Li6{^{6}\rm{Li}}                       

\newcommand{\ud}{\mathrm{d}}

\makeatother

\begin{document}
	
	\title{Median-point approximation and its application for the study of fermionic systems}
	
	\author{D.~Makogon}
	\noaffiliation
	\author{C. Morais Smith}
	\affiliation{Institute for Theoretical Physics, Utrecht University,
		Princetonplein 5, 3584 CC Utrecht, The Netherlands}
	
	\date{\today}
	
	\begin{abstract}
		We consider a system of fermions with local interactions on a lattice (Hubbard model) and apply a novel extension of the Laplace's method (saddle-point approximation) for evaluating the corresponding partition function. There, we introduce dual free bosonic fields, with a propagator corresponding to an effective (renormalized) interaction with Maki-Thompson and Aslamazov-Larkin type corrections and beyond, and demonstrate that the superconducting pairing originates as an instability of the effective interaction. We derive the corresponding Bethe-Salpeter equation (instability criterion) and show that the interaction enters the equation only in its effective form to all orders, including the exchange part of the self-energy. An important implication of this result is that the effective interaction always remains finite, even at phase-transition points, directly contradicting the often used assumption of linear relationship between the interaction and susceptibility, established within the random-phase approximation. By analyzing the Bethe-Salpeter equation in the context of unconventional superconductivity, we find that the presence of a flat band close the Fermi level, found in materials such as twisted bilayer graphene, has a twofold favorable impact persisting beyond the weak-coupling approximation: a reduced kinetic energy cost of the gap formation and an increased anisotropy of the effective interaction, favoring a momentum dependent order parameter.   
	\end{abstract}

	
	\maketitle
	
	\section{Introduction}
	Strongly correlated systems are at the core of condensed-matter physics. Instability towards the formation of a superconducting state, a charge-density wave, or a magnetic texture, such as ferromagnetic, anti-ferromagnetic or more general spin-density waves are among the first topics to be investigated in any newly discovered material.\cite{Cao1, Cao2, GuohongLi, Pixley} The theoretical treatment of these phenomena usually rely on Hartree-Fock (HF) or random-phase approximations (RPA),\cite{Bohm_Pines} and are restricted to the particle-particle or particle-hole channels. In this way, a mean-field approximation in the particle-particle channel can successfully describe superconductivity in many materials, whereas in the particle-hole channel it accounts for spin-density or charge-density waves, depending on the sign of the interactions.\cite{Peres} This separation of the problem into particle-particle or particle-hole channels hampers a mean-field description of systems in which superconductivity and charge-density wave formation occur concomitantly, such as in Transition-Metal Dichalcogenides (TMD)\cite{Wilson_DiSalvo_Mahajan,McMillan,Machida} and in high-Tc superconductors.\cite{Varma_Littlewood_Schmitt-Rink_Abrahams_Ruckenstei,Gabovich,Chang}  Generalized phenomenological formulations including both phenomena on equal footing were proposed, such as an SO(5) model.\cite{Zhang} 
	
	It has been long recognized that even proximity to a magnetically ordered state can have strong impact on a superconducting phase and pairing interaction in general. Namely, using a diagrammatic re-summation technique, it was shown\cite{Berk_Schrieffer, Doniach_Engelsberg} that paramagnon exchange in the vicinity of a ferromagnetic phase transition strongly enhances the repulsive interaction, therefore, suppressing pairing. However, later it was proposed that spin fluctuation exchange can actually provide a pairing mechanism for $p$-wave $^3$He superfluidity,\cite{Layzer_Fay} while the idea of anisotropic solutions to the BCS equation itself, mostly in the context of $^3$He superfluidity, has emerged earlier.\cite{Pitaevskii,Thouless,Brueckner_Soda_Anderson_Morel} Subsequently, spin pairing was proposed also in the context of organic superconducting materials, \cite{Emery} as well as for the $d$-wave superconductivity mechanism for some heavy fermions and high-Tc superconductors \cite{Cyrot, Scalapino_Loh_Hirsch_1986, Miyake_1986}(a more detailed overview is given in Ref.\;[\onlinecite{Scalapino_overview}]).
	
	One of the simplest models proposed to capture the mechanism of high-Tc superconductivity is the Hubard model, originally introduced to describe electrons in solids.\cite{Hubbard_1963} A rather convincing evidence for the existence of a superconducting phase in the Hubbard model has been provided in Ref.\;[\onlinecite{Metzner}] using the functional renormalization group (FRG) approach, where the RG flow of two and four -point vertex functions is evaluated numerically with a number of approximations. The onset of pairing instability is evidenced by the appearance of a pronounced attractive interaction with zero total inbound momentum, subsequently leading to a superconducting state with finite order parameter. \cite{Eberlein_Metzner} However, FRG is foremost a numerical procedure,
	and deriving an analytical expression for the instability criterion is not straightforward. It is also not directly evident why specifically a $d$-wave gap emerges. These can be intuitively understood already with the original  expression for the effective interaction in the BCS gap equation from Refs.\; [\onlinecite{Berk_Schrieffer}, \onlinecite{Doniach_Engelsberg}], demonstrating that in the vicinity of an anti-ferromagnetic phase transition, the effective interaction becomes highly anisotropic, allowing for a $d$-wave gap solution. This result is also supported by Monte Carlo simulation.\cite{White_Scalapino} As a next step, a semi-analytical fluctuation-exchange (FLEX) approximation was proposed for the evaluation of one- and two-particle correlation functions. \cite{Bickers_Scalapino_White,Bickers_Scalapino} The FLEX approximation takes into account corrections to the self-energy beyond the Hartree-Fock approximation, which are given by the sum of T-matrix ladder diagrams in the $pp$- and $ph$- channels, while subtracting certain terms to prevent double counting. \cite{Esirgen_Bickers} The corrections are linear in the "fluctuation propagator", which is essentially the effective interaction $W$ in the random-phase approximation for the particle-hole channel, and similarly for the particle-particle channel. The pairing instability is described by the Bethe-Salpeter equation\cite{Nambu, Salpeter_Bethe}, with the effective interaction in RPA approximation. Another often applied many-body perturbation theory (MBPT) is the $GW$ approximation, introduced in Refs.\;[\onlinecite{Hedin_1965}, \onlinecite{Hedin_Lundqvist_1971}], which has been successfully applied for an {\it ab initio} calculation of the band structure of real materials. There, the exchange part of the self-energy is calculated with the effective interaction $W$, instead of the bare one, as in the Hartree-Fock approximation. Besides, a number of works\cite{Dahm_Scalapino,Esirgen_Schuttler_Grober_Evertz} exploit the linear relationship between the effective interaction and the susceptibility established within RPA\cite{Anderson_Brinkman_1973} or suppose the validity of such a dependence. \cite{Abanov_Chubukov_Schmalian,Chubukov_Pines_Schmalian} Clearly, this choice implies a diverging effective interaction, with a diverging susceptibility at the point of the phase transition, and thus, prevents an adequate description of pairing and of the transition itself with perturbative approaches. This problem can be related to the failure of FLEX (and other MBPTs at RPA level) to outperform second-order perturbation results for the Hubbard model close to half filling.\cite{Gukelberger_Huang_Werner} Moreover, the linear relationship was used to establish a link between the exact charge susceptibility, estimated via Monte Carlo simulation techniques, and the supposedly exact effective interaction.\cite{Esirgen_Schuttler_Grober_Evertz}
	
	Here, we develop a different  approach, based on the use of auxiliary Hubbard-Stratonovich fields coupled to fermionic spin and charge densities. As a main outcome, we give an exact definition of the effective interaction $W$, demonstrating that the linear relationship between the effective interaction and the susceptibility is an artifact of RPA, which does not hold in general, when higher-order interaction terms are taken into account. We also show that these two quantities do not necessarily diverge simultaneously. Moreover, we demonstrate that the effective interaction $W$ should remain finite even at a phase transition point. We concentrate on fermionic systems with local interactions, which undergo a second-order phase transition and are well described by a Landau free energy\cite{Landau} for a suitably chosen order parameter. The free energy can be found by performing a Legendre transformation of the corresponding effective action. Consequently, the Taylor series expansion of the free energy in the order parameter is related to the corresponding correlation functions, while the phase transition should be signaled by a diverging correlation function. These can be found as derivatives of the corresponding partition function with respect to generic source terms. In particular, the possibility of a second-order phase transition into a superconducting state can be investigated when the partition function is known with sufficient accuracy. Because, in general, the partition function can not be evaluated exactly, the main challenge is the construction of an accurate approximation. For example, the partition function is often evaluated within the saddle-point approximation after performing a Hubbard-Stratonovich transformation\cite{Stratonovich, Hubbard_1959} and integrating out the fermionic fields. Currently, the saddle-point approximation can either reproduce the RPA results, when applied within the particle-hole channel, or the Cooper instability,\cite{BardeenCooper} when applied within the particle-particle channel. A naive combination of the contributions of the two channels results into over-counting of diagrams and multiple efforts have been made to remedy this problem.\cite{Scuseria, Tahir} In particular, a number of counter-terms are introduced in FLEX in order to prevent over-counting at a given expansion order. However, no consistent and systematic approach has been presented up to now.
	Our approach is based on the observation that the Hubbard-Stratonovich transformation is exact, even when applied only within the particle-hole channel. Therefore, a generalization of both results involves higher-order terms beyond the saddle-point approximation. It has been shown in Refs.\;[\onlinecite{Hertz}, \onlinecite{Millis}] that expanding the action up to quartic order in the bosonic fields and applying the RG technique leads in many cases to a correct temperature dependence of the correlation length near the quantum critical point. However, below we demonstrate that the contribution of any finite number of terms in the particle-hole channel does not lead to the pairing instability, which emerges only when the contribution from an {\it infinite} number of terms is taken into account.
	
	Considering a system of fermions with local interactions on a lattice and performing a Hubbard-Stratonovich transformation within the particle-hole channel, we arrive at the equivalent bosonic action after integrating out the fermionic fields. In order to evaluate the corresponding bosonic partition function, we consider a change of integration variables that makes the action strictly quadratic.  Matching of all derivatives of the action yields an infinite system of nonlinear equations, describing the mapping to the corresponding quadratic bosonic action, i.e. free boson distribution. The distribution is characterized by the median point, $\mathbf{M}_{\rm sp}$, interpreted as an effective field at which the fermionic propagators, $\mathbf{G}(\mathbf{M})$, are evaluated and the bosonic propagator, $\mathbf{W}$, is interpreted as the effective interaction. The definition implies that the partition function $Z$ is proportional to $\det[\mathbf{W}]^{1/2}$. Therefore, as long as the partition function remains finite, which holds at the point of the phase transition, the effective interaction $\mathbf{W}$ also does (namely, $\det[\mathbf{W}]$, assuming a finite $\mathbf{M}_{\rm sp}$). The corresponding exact expressions for the effective field and interaction have the form of the saddle-point approximation values, with additional terms originating from the Jacobian of the transformation. In particular, the effective field $\mathbf{M}_{\rm sp}$ acquires a correction term proportional to the second derivative of the transformation, while the effective interaction inverse $\mathbf{W}^{-1}$ acquires correction terms proportional to the third and to the squared second derivatives of the transformation. We identify that the former gives rise to Maki-Thompson\cite{Maki,Thompson}  and the latter to Aslamazov-Larkin\cite{Aslamazov_Larkin}  type corrections to the effective interaction. In order to arrive at a mathematically controllable approximation scheme ("median-point approximation"), we consider the action with a generalized fermion propagator in the form of the weighted geometric mean of the original propagator $\mathbf{G}(\mathbf{M})$ and the one evaluated at the effective field $\mathbf{G}(\mathbf{M}_{\rm sp})$. The saddle-point approximation corresponds to the case with zero weight for the original propagator, while the exact form is recovered when the weight is one. A solution of the system of equations up to the leading-order approximation in the weight corresponds to a sum of diagrams with one fermionic loop. Below, we demonstrate that the expansion of the effective field and the interaction inverse  $\mathbf{W}^{-1}$ in orders of the weight corresponds to a resummation of diagrams, with the number of fermionic loops given by the order. 
	
	The chosen Hubbard-Stratonovich transformation within the particle-hole channel, with the auxiliary fields coupled to fermionic spin and charge densities, is well suited for a description of the normal state in the presence of different instabilities, such as spin density wave (SDW) or charge density wave (CDW), and their onset is signaled by an instability of the effective field $\mathbf{M}_{\rm sp}$.  However,  the approach is not suitable for a treatment of the superconducting state with a finite superconducting gap parameter, which can be included only as an infinitesimal perturbation. In fact, investigating the stability of the normal state with respect to such perturbations is sufficient to identify the superconducting pairing instability. By analyzing the stability of the effective interaction to the presence of an infinitesimally small pairing interaction within the leading-order approximation, we establish the equivalence of a particular class of resulting higher-order terms with a 'wheel' structure diagrams in the particle-hole channel to the 'ladder' diagrams in the particle-particle channel, the so-called $pp$-ladders. The resummation of these diagrams leads to the Bethe-Salpeter equation (approximated to the first order in the effective interaction) and thus, the Cooper instability condition in a particular limiting case. Therefore, we demonstrate the emergence of a pairing instability (leading to the superconducting state) as an interaction instability in the particle-hole channel, without explicitly introducing the particle-particle channel. This is in line with the FRG results,\cite{Eberlein_Metzner}   but somewhat contradicts the reported superconducting instability driven by diverging couplings describing the effective interaction between the "hot spots" in the square lattice,\cite{Furukawa} in graphene,\cite{Nandkishore} and in twisted bilayer graphene\cite{Isobe}. The inconsistency can be attributed to the differences in the coupling definitions, making them not suitable for direct comparison, or to an artifact of the "hot spots" approximation, as the RG flow in the lowest coupling order is known for spurious divergencies, such as the Landau pole.\cite{Landau_pole}

Resummation of more general ladder diagrams (the so-called "X-ladders",\cite{Gukelberger_Huang_Werner} which include not only $ph$- and $pp$-ladders, but also diagrams with crossed interaction lines) leads to the appearance of terms with higher order in the effective interaction in the Bethe-Salpeter equation. The resulting equation is analogous to the one originally derived in Refs.\;[\onlinecite{Berk_Schrieffer} and  \onlinecite{Doniach_Engelsberg}], but with the effective interaction $W$, instead of the bare Hubbard $U$. Moreover, we demonstrate that the interaction enters the Bethe-Salpeter equation only in its effective form $W$ to all orders, including the exchange part of the self-energy, analogous to the $GW$ approximation.\cite{Hedin_1965} Here, $W$ is the effective interaction with spin- and charge- density fluctuations, which is relevant in the context of unconventional superconductors.\cite{Scalapino_Loh_Hirsch_1986, Vuletic, BenAli, Manske, Raghu}  Besides the renormalized propagator $G$ and interaction $W$, the vertex function is also renormalized in the Bethe-Salpeter equation. The three quantities are described by a system of self-consistent equations, which we write down at one-loop level, excluding cross term diagrams. Considering a hypothetical situation when $W$ is known exactly (or can be determined numerically with sufficient precision) and diagrams of all orders are present, the Bethe-Salpeter equation in this form provides an approximation, re-summing the simplest types of diagrams, with renormalized vertex and fermionic propagator (contrary to the expression for $W$, which involves only a "bare"  propagator, without the exchange part of the self-energy). Clearly, the approximation becomes inaccurate when the omitted terms provide a comparable contribution, as it happens close to the point of instability. Nevertheless, the equations provide a systematic improvement over the mean-field theory results and allow for the study of the pairing instability in the presence of SDW or CDW. Moreover, we demonstrate that solving perturbatively the system of the self-consistent equations in quantum electrodynamics (QED) is analogous to the renormalization procedure (at least, at one loop level), with the advantage that no counter terms have to be explicitly introduced. Therefore, in the context of QED, the effective interaction $W$ has the same interpretation as a renormalized (finite) interaction coupling combined with the renormalized photon propagator.
The name "median-point approximation" stems from the fact that the saddle-point of the Gaussian distribution coincides with the median-point, which remains median upon change of integration variables that might move the saddle-point.

The outline of the paper is the following. We start by presenting a generic formulation of the problem in Sec. II. Then, we first reproduce the saddle-point approximation results in Sec. IIIA, before introducing our approximation scheme in Sec. IIIB. In Sec. IV, we show how the system of self-consistent equations derived within this procedure is related to correlation functions. Finally, in Sec. V we demonstrate the possibility of describing a superconducting phase transition and derive the corresponding criterion. Our conclusions are presented in Sec. VI. 

\section{Lattice models with local interactions}
Let us consider a generic tight-binding model for spin-$1/2$ fermions in a lattice with arbitrary geometry, described by a Hamiltonian $H = H_0 + H_{\rm int}$, where the tight-binding term reads  
\begin{equation}
\hat{H}_{0}=-t\sum_{<\mathbf{i},\mathbf{j}>,s}c^{\dag}_{\mathbf{i},s}c_{\mathbf{j},s} + {\rm h.c. }, 
\end{equation} 
with $t$ denoting the hopping parameter and $c^{(\dagger)}$ the annihilation (creation) operators, and local interactions, 
\begin{equation}\label{H_Hubbard}
\hat{H}_{\rm
int}=U\sum_{\mathbf{j}}c^{\dag}_{\mathbf{j},\uparrow}c^{\dag}_{\mathbf{j},\downarrow}c_{\mathbf{j},\downarrow}c_{\mathbf{j},\uparrow},
\end{equation}
where $U$ is the Hubbard on-site interaction. \cite{Hubbard_1963}

Our aim is to determine whether a second-order
phase transition occurs in this system due to the effect of local
fermion-fermion interactions. Usually, in 1D systems a
charge-density wave transition arises due to the coupling of the
fermionic degrees of freedom with the phonons of the lattice. This
coupling leads to an effective attractive interaction between the
fermions, which combined with the nesting of the wavevector in the
particular Fermi surface, results into a dimerization of the ions
and a consequent electronic charge-density-wave instability. Phase
transitions can be well described by the Landau theory, which
includes an expansion of the free energy in terms of the order
parameter around the critical temperature at which the transition
occurs. Here, we investigate the possibility of occurrence of a second-order phase transition,
which can be determined by the point at which the coefficient of the
second-order term in the Landau free energy vanishes. 

By introducing the coherent states formalism, we may write the
grand-canonical partition function as
\begin{equation}
Z=\int \mathrm{d}[c^{\dag}]\mathrm{d}[c]e^{-{\cal
S}[c^{\dag},c]/\hbar},
\end{equation}
where the Euclidean action is
\begin{equation}
{\cal S}=\int_0^{\hbar\beta} \mathrm{d} \tau \left[\sum_{\mathbf{j}}
c^{\dag}_{\mathbf{j}} \left(\hbar\frac{\partial}{\partial
\tau}-\mu\right)c_{\mathbf{j}}+H_0+H_{\rm int}\right],
\end{equation}
with $\beta=1/k_BT$, $\tau$ the imaginary time variable and $\mu$ the chemical potential. \\

\subsection{Generic formulation}

It is convenient to rewrite the interaction
term in the quadratic form 
\begin{equation}\label{intterm_sz}
c^{\dag}_{\mathbf{j},\uparrow}c^{\dag}_{\mathbf{j},\downarrow}c_{\mathbf{j},\downarrow}c_{\mathbf{j},\uparrow}=\frac{1}{4}n_\mathbf{j}^2-S_{\mathbf{j},z}^2,
\end{equation}
with $n_\mathbf{j}=c_{\mathbf{j}}^{\dag}c_{\mathbf{j}}$ and $S_{\mathbf{j},z}=c_{\mathbf{j}}^{\dag}\sigma_3 c_{\mathbf{j}}/2$. The
operator product $c^{\dag}c$ (and analogously $c^{\dag}\sigma_3 c$) denotes
$c^{\dag}(\tau+\zeta)c(\tau)$, with $\zeta\rightarrow0+$. The fact that the spin and charge terms in
Eq.\;(\ref{intterm_sz}) enter with a different sign reflects the Pauli
principle, which implies a vanishing self-energy for a polarized
state in the case of only local interactions. 
As a next step, we perform a Hubbard-Stratonovich transformation, which
renders the action quadratic in the fermion operators, at the cost of
introducing auxiliary bosonic fields. The form in Eq.\;(\ref{intterm_sz}) 
explicitly introduces a preferred direction along the $z$-axis, while the interaction term has rotational symmetry and remains invariant with respect to an $SO(3)$ transformation.  Therefore, different choices of transformation are possible, depending on the form of the interaction term. For example, another equivalent form is 
\begin{equation}\label{intterm}
	c^{\dag}_{\mathbf{j},\uparrow}c^{\dag}_{\mathbf{j},\downarrow}c_{\mathbf{j},\downarrow}c_{\mathbf{j},\uparrow}=\frac{1}{4}n_\mathbf{j}^2-{\mathbf{S}_\mathbf{j}}\cdot\mathbf{m}_\mathbf{j}\cdot\mathbf{S}_\mathbf{j},
\end{equation}
where $\mathbf{m}_\mathbf{j}$ is a symmetric matrix with ${\rm
Tr}[\mathbf{m} _\mathbf{j}]= 1$ and  $\mathbf{S}_\mathbf{j}=c_{\mathbf{j}}^{\dag}\sigma c_{\mathbf{j}}/2$. Although all transformations are exact, they lead to somewhat different approximations and thus, different free energy estimates. One possibility to resolve this ambiguity is to select the form that minimizes the ground-state energy estimate or free energy, according to the "principle of minimal sensitivity".\cite{Stevenson_1982} In particular, if only the $z$-axis is relevant, the form in Eq.\;(\ref{intterm_sz}) becomes the most appropriate (also making it convenient to introduce the particle-particle channel, representing fermions via Nambu spinors). Here, we do not follow this approach, but only consider the generic form given by Eq.\;(\ref{intterm}) with a certain $\mathbf{m}_\mathbf{j}=\mathbf{m}$, unless stated otherwise. Without loss of generality, we assume $\mathbf{m}$ to be invertible, as degenerate cases can be obtained in an appropriate limit. For illustration, we consider a lattice geometry with two sublattices $A$ and $B$, and denote by $a$ and $b$ the corresponding fermionic operators. In this case, the auxiliary bosonic fields are
 $\phi^a$, $\phi^b$,
$\mathbf{M}^a$, and $\mathbf{M}^b$, where $\mathbf{M}^a$ and
$\mathbf{M}^b$ are three component vectors. The choice of the
Hubbard-Stratonovich transformation allows us to consider ground
states with or without spontaneous symmetry breaking in the
particle-hole channel, which are realized by introducing
infinitesimal symmetry breaking fields, while searching for an
instability to a superconducting state. We define the path integral
measures $\mathrm{d}[\phi^a]$ and $\mathrm{d}[\mathbf{M}^a]$ such
that
\begin{widetext}
\begin{equation}
e^{-\frac{U}{\hbar}\int_0^{\hbar\beta} \mathrm{d} \tau
\sum_{\mathbf{j}\in
A}a^{\dag}_{\mathbf{j},\uparrow}a^{\dag}_{\mathbf{j},\downarrow}a_{\mathbf{j},\downarrow}a_{\mathbf{j},\uparrow}}=\int
\mathrm{d}[\phi^a]\mathrm{d}[\mathbf{M}^a]e^{-\frac{1}{\hbar U
}\int_0^{\hbar\beta} \mathrm{d} \tau \sum_{\mathbf{j}\in
A}\left[{(\mathbf{M}^a_\mathbf{j}\cdot\mathbf{m}^{-1}-2 U \mathbf{S}_\mathbf{j})} \cdot
\mathbf{M}^a_\mathbf{j}-\phi^a_\mathbf{j}(\phi^a_\mathbf{j}- U
n^a_\mathbf{j})\right]}
\end{equation}
for the $A$ sublattice, and analogously for the $B$ sublattice.
Notice that the integration for $\phi^a$ and $\phi^b$ goes along a
contour parallel to the imaginary axis from $-i\infty$ to
$+i\infty$ with a certain offset. The partition function after the transformation is
\begin{equation}
Z=\int
\mathrm{d}[a^{\dag}]\mathrm{d}[a]\mathrm{d}[b^{\dag}]\mathrm{d}[b]\mathrm{d}[\phi^a]
\mathrm{d}[\mathbf{M}^a]\mathrm{d}[\phi^b]\mathrm{d}[\mathbf{M}^b]e^{-({\cal
S}_0+{\cal S}_1+{\cal S}_2)/\hbar},
\end{equation}
where the action
\begin{equation}
{\cal S}_0=\int_0^{\hbar\beta} \mathrm{d} \tau \left[
\sum_{\mathbf{j}\in A} a^{\dag}_{\mathbf{j}}
\left(\hbar\frac{\partial}{\partial
\tau}-\mu\right)a_{\mathbf{j}}+\sum_{\mathbf{j}\in B}
b^{\dag}_{\mathbf{j}} \left(\hbar\frac{\partial}{\partial
\tau}-\mu\right)b_{\mathbf{j}}+H_0\right],
\end{equation}
\begin{equation}
{\cal S}_1=-\int_0^{\hbar\beta} \mathrm{d} \tau
\left[\sum_{\mathbf{j}\in
A}(2\mathbf{S}_\mathbf{j}\cdot\mathbf{M}^a_\mathbf{j}-n^a_\mathbf{j}\phi^a_\mathbf{j})
+\sum_{\mathbf{j}\in
B}(2\mathbf{S}_\mathbf{j}\cdot\mathbf{M}^b_\mathbf{j}-n^b_\mathbf{j}\phi^b_\mathbf{j})\right],
\end{equation}
and
\begin{equation}
{\cal S}_2=\frac{1}{ U }\int_0^{\hbar\beta} \mathrm{d} \tau
\left\{\sum_{\mathbf{j}\in A}[\mathbf{M}^a_\mathbf{j}\cdot\mathbf{m}^{-1} \cdot
\mathbf{M}^a_\mathbf{j}
-(\phi^a_\mathbf{j})^2]+\sum_{\mathbf{j}\in
B}[\mathbf{M}^b_\mathbf{j}\cdot\mathbf{m}^{-1} \cdot
\mathbf{M}^b_\mathbf{j}-(\phi^b_\mathbf{j})^2]\right\}.
\end{equation}

Defining the Fourier transformations
\begin{equation}
a_{\mathbf{j}}(\tau)=\frac{1}{\sqrt{N\hbar\beta}}\sum_{\omega_n,\mathbf{k}}a_{\omega_n,\mathbf{k}}e^{-i\omega_n\tau+i
\mathbf{k} \cdot \mathbf{j}}
\end{equation}
and
\begin{equation}
\mathbf{M}^a_{\mathbf{j}}(\tau)=\sum_{\Omega_n,\mathbf{k}}\mathbf{M}^a_{\Omega_n,\mathbf{k}}e^{-i\Omega_n\tau+i
\mathbf{k} \cdot \mathbf{j}},
\end{equation}
where the fermionic Matsubara frequency is 
$\omega_n=\pi(2n+1)/\hbar\beta$, the bosonic Matsubara frequency is $\Omega_n=2\pi n/\hbar\beta$, and the number of sites for each sublattice is $N$, allows us to rewrite the action as
\begin{equation}
{\cal
S}=-\hbar\sum_{\mathbf{q},\mathbf{q}'}\psi^\dag_{\mathbf{q}}
\cdot (\mathbf{G}_{0\;\mathbf{q},\mathbf{q}'}^{-1}+\mathbf{C}_{\mathbf{q},\mathbf{q}'}) \cdot \psi_{\mathbf{q}'} 
+{\cal S}_2,
\end{equation}
where $\mathbf{q} = (\omega_n,\mathbf{k})$ is a combined frequency-momentum vector and we denote
\begin{equation}
\psi_{\mathbf{q}}\equiv\left( \begin{array}{c} a_{\mathbf{q}}\\
b_{\mathbf{q}}\nonumber \end{array} \right).
\end{equation} 

The bare inverse Green's function is
\begin{equation}
-\hbar\mathbf{G}_{0\;(\omega_n,\mathbf{k}),(\omega_{n'},\mathbf{k}')}^{-1}=\mathbf{H}_\mathbf{k}\delta_{\mathbf{k},\mathbf{k}'}\delta_{n,n'}+ \left[
\begin{array}{cc}  -(\mu
+i\hbar\omega_n)\mathbf{I} & 0\\
0 & -(\mu +i\hbar\omega_n)\mathbf{I} \nonumber
\end{array} \right]\delta_{\mathbf{k},\mathbf{k}'}\delta_{n,n'},
\end{equation}
where $\mathbf{I}$ denotes the $2\times2$ identity matrix and
$\delta_{\alpha,\beta}$ is the Kronecker delta. 
In particular, for 
\begin{equation}
\mathbf{H}_\mathbf{k}=\left[
\begin{array}{cc} 0  & -t\gamma^*_\mathbf{k}\mathbf{I}\\
-t\gamma_\mathbf{k}\mathbf{I} & 0 \nonumber
\end{array} \right],
\end{equation}
with $\gamma_\mathbf{k}$ denoting the sum of exponents over the nearest neighbors, $\gamma_\mathbf{k}\equiv \sum_{\mathbf{j}}e^{i \mathbf{k} \cdot \mathbf{j}},$
the bare inverse Green's function reads 
\begin{equation}
-\hbar\mathbf{G}_{0\;(\omega_n,\mathbf{k}),(\omega_{n'},\mathbf{k}')}^{-1}=\left[
\begin{array}{cc} -(\mu
+i\hbar\omega_n)\mathbf{I} & -t\gamma^*_\mathbf{k}\mathbf{I}\\
-t\gamma_\mathbf{k}\mathbf{I} & 
-(\mu +i\hbar\omega_n)\mathbf{I} \nonumber
\end{array} \right]\delta_{\mathbf{k},\mathbf{k}'}\delta_{n,n'}.
\end{equation}
Because the
transformation from $\mathbf{M}^a_{\mathbf{j}}(\tau)$ to
$\mathbf{M}^a_{\omega_n,\mathbf{k}}$ is not unitary, the
path-integral measure is adjusted by a constant factor. The coupling with a fermionic field is given by
\begin{equation}
\hbar\mathbf{C}_{(\omega_n,\mathbf{k}),(\omega_{n'},\mathbf{k}')}=e^{i\zeta
\omega_n }\left[
\begin{array}{cc} \mathbf{\sigma} \cdot
\mathbf{M}^a_{\omega_{n}-\omega_{n'},\mathbf{k}-\mathbf{k}'}-\phi^a_{\omega_{n}-\omega_{n'},\mathbf{k}-\mathbf{k}'}\mathbf{I}& 0\\
0 & \mathbf{\sigma} \cdot
\mathbf{M}^b_{\omega_{n}-\omega_{n'},\mathbf{k}-\mathbf{k}'}-\phi^b_{\omega_{n}-\omega_{n'},\mathbf{k}-\mathbf{k}'}\mathbf{I}\nonumber
\end{array} \right].
\end{equation}
The coupling term is linear in the bosonic fields. Therefore, a more generic form is 
\begin{equation}
\mathbf{C}_{\mathbf{q},\mathbf{q}'}=\sum_{\mathbf{p},r}\mathbf{R}_{\mathbf{q},\mathbf{q}';\mathbf{p},r}M^{(\mathbf{p},r)}
\end{equation}
where $r$ indicates a component of the eight-component vector $\mathbf{M}^{\mathbf{p}}$, which is defined as
\begin{equation}
\mathbf{M}^{\mathbf{p}}=\left[
\begin{array}{c}\phi^a_{\mathbf{p}}\\
\mathbf{M}^a_{\mathbf{p}}\\
\phi^b_{\mathbf{p}}
 \\ \mathbf{M}^b_{\mathbf{p}}\end{array} \right]
\end{equation}
and the tensor $\mathbf{R}$ is
\begin{equation}\label{P_matrices}
\mathbf{R}_{(\omega_n,\mathbf{k}),(\omega_{n'},\mathbf{k}');(\Omega_m,\mathbf{K}),r}=\frac{e^{i\zeta
\omega_n }}{\hbar}\sum_{r'}\mathbf{P}^{r'} \eta_{r'r}
\delta_{n-n',m}\delta_{\mathbf{k}-\mathbf{k}',\mathbf{K}},
\end{equation}
where $\mathbf{P}^{r'}$ are constant $4\times4$ matrices and the
exponential prefactor originates from the earlier mentioned operator
ordering, with $\zeta\rightarrow0+$. The matrix $\eta={\rm
Diag}(-1,1,1,1,-1,1,1,1)$
 is the metric signature. 
In a shorthand notation, with $\mathbf{P}$ being a vector composed
of $\mathbf{P}^r$ matrices, $\mathbf{P}=[{\rm
Diag}(\mathbf{I},0)$, ${\rm Diag}(\sigma_1,0)$, ${\rm
Diag}(\sigma_2,0)$, ${\rm Diag}(\sigma_3,0)$, ${\rm
Diag}(0,\mathbf{I})$, ${\rm Diag}(0,\sigma_1)$, ${\rm
Diag}(0,\sigma_2)$, ${\rm Diag}(0,\sigma_3)]^T$. A different choice of the Hubbard-Stratonovich transformation would lead to a different expression for the vector $\mathbf{P}$, but this has no impact on the generic formalism presented below.

Now, we introduce the inverse Green's function as
\begin{equation}\label{Greens_function_with_h}
\mathbf{G}_{\mathbf{q},\mathbf{q}'}^{-1}(\mathbf{M})=\mathbf{G}_{0\;\mathbf{q},\mathbf{q}'}^{-1} + \sum_{\mathbf{p},r}\mathbf{R}_{\mathbf{q},\mathbf{q}';\mathbf{p},r}M^{(\mathbf{p},r)}+ \mathbf{h}_{\mathbf{q},\mathbf{q}'},
\end{equation}
where a generating field $\mathbf{h}$ is added. Omitting the indices and assuming the summation convention, the equation reads 
\begin{equation}\nonumber
\mathbf{G}^{-1}(\mathbf{M})=\mathbf{G}_{0}^{-1}+ \mathbf{R}\mathbf{M}+ \mathbf{h}.
\end{equation}
Here, $\mathbf{M}$ can be thought of as a generic bosonic (auxiliary) field that couples linearly to the fermionic fields. Clearly, 
\begin{equation}\nonumber
\ud\mathbf{G}^{-1}(\mathbf{M})= \mathbf{R}\ud\mathbf{M},
\end{equation}
and therefore, the tensor $\mathbf{R}$ can be interpreted as the derivative of $\mathbf{G}^{-1}(\mathbf{M})$ with respect to $\mathbf{M}$.
In this notation
\begin{equation}
 {\cal S}_2 =\frac{N\hbar\beta}{ U
}\sum_{\mathbf{p},r',r}M^{(\mathbf{p},r)}(\mathbf{g}^{-1})_{r,r'}
M^{(-\mathbf{p},r')},
\end{equation}
where $\mathbf{g}={\rm	Diag}(-1,\mathbf{m},-1,\mathbf{m})$. For convenience, we introduce the symmetric covariant tensor
$\mathbf{U}^{-1}_{(\mathbf{p},r),(\mathbf{p}',r')}=2U^{-1}\mathbf{g}^{-1}
\delta_{\mathbf{p},-\mathbf{p}'}$ acting on the contravariant vectors $\mathbf{M}$ and omit indices in the following
tensorial notation
\begin{equation}
\sum_{(\mathbf{p},r),(\mathbf{p}',r')}(\mathbf{U}^{-1})_{(\mathbf{p},r),(\mathbf{p}',r')}M^{(\mathbf{p},r)}
M^{(\mathbf{p}',r')}=\mathbf{U}^{-1}\mathbf{M} \mathbf{M}.
\end{equation}

The partition function in the generic notation reads
\begin{equation}\label{Z_psi_M}
Z[\mathbf{h}]=\int \mathrm{d}[\psi^{\dag}]\mathrm{d}[\psi]\mathrm{d}[\mathbf{M}]\exp\left(\psi^{\dag}\mathbf{G}^{-1}(\mathbf{M})\psi-\frac{N\beta}{ 2 }\mathbf{U}^{-1}\mathbf{M}\mathbf{M}\right).
\end{equation}
Moreover, we can consider even the more general case of non-local interactions, when $\mathbf{M}$ and $\mathbf{U}$ depend on the inbound momentum $\mathbf{q}'$,  
\begin{equation}
\mathbf{G}_{\mathbf{q},\mathbf{q}'}^{-1}(\mathbf{M})=\mathbf{G}_{0\;\mathbf{q},\mathbf{q}'}^{-1} + \sum_{\mathbf{p},r}\mathbf{R}_{\mathbf{q},\mathbf{q}';\mathbf{p},r}M^{(\mathbf{p},r)}_{\mathbf{q}'}+ \mathbf{h}_{\mathbf{q},\mathbf{q}'}
\end{equation}
and
\begin{equation}
\sum_{(\mathbf{p},r),(\mathbf{p}',r'),\mathbf{q},\mathbf{q}'}(\mathbf{U}^{-1})^{\mathbf{q},\mathbf{q}'}_{(\mathbf{p},r),(\mathbf{p}',r')}M^{(\mathbf{p},r)}_{\mathbf{q}}
M^{(\mathbf{p}',r')}_{\mathbf{q}'}=\mathbf{U}^{-1}\mathbf{M} \mathbf{M}.
\end{equation}
Specifically, the interaction might depend on the total momentum $\mathbf{q}+\mathbf{q}'$
\begin{equation}\label{u_total_inbound}
\sum_{(\mathbf{p},r),(\mathbf{p}',r'),\mathbf{q},\mathbf{q}'}(\mathbf{U}^{-1})^{\mathbf{q}+\mathbf{q}'}_{(\mathbf{p},r),(\mathbf{p}',r')}M^{(\mathbf{p},r)}_{\mathbf{q}}
M^{(\mathbf{p}',r')}_{\mathbf{q}'}.
\end{equation}

In what follows, unless explicitly indicated, we consider interactions which do not dependent on the inbound momentum. 
\subsection{Generic interactions beyond on-site}
In the previous section, we have considered only an on-site interaction of the form of Eq.\;(\ref{H_Hubbard}). A more generic interaction term with a non-zero off-site component is
\begin{equation}
	\hat{H}_{\rm
		int}=\frac{1}{2}\sum_{\mathbf{i},\mathbf{j}}V_{\mathbf{i}\mathbf{j}}\hat{n}_\mathbf{i}\hat{n}_\mathbf{j},
\end{equation}
which provides a more realistic approximation of the (screened) electron-electron interaction $V(\mathbf{r} -\mathbf{r'})$ related by\cite{Goodwin} 
\begin{equation}
	V_{\mathbf{i}\mathbf{j}}=\iint d \mathbf{r} d\mathbf{r'}|w_\mathbf{i}(\mathbf{r'} )|^2 V(\mathbf{r} -\mathbf{r'})|w_\mathbf{j}(\mathbf{r} )|^2,
\end{equation} 
with the Wannier functions $w_\mathbf{i}(\mathbf{r})$ for the lattice of the material, such as graphene\cite{Jung} or Twisted Bilayer Graphene (TBG).\cite{Carr,Goodwin_R, Koshino, Kang} 
A number of works\cite{Pizarro,Goodwin,Samajdar_Scheurer,Cea, Cea3,Dai,Dong}  have proposed superconductivity in Twisted Bilayer, Trilayer, and double Bilayer Graphene to be caused by an anisotropy of the effective electron interaction (screened Coulomb interaction) that enters the gap equation. However,  some of the approaches\cite{Pizarro,Goodwin,Samajdar_Scheurer,Cea, Cea3} consider only dielectric screening at RPA level in the charge density channel, without the influence of spin fluctuations, which are expected to provide a significant impact\cite{Dai,Dong} in the vicinity of a spin ordered phase. Below, we demonstrate how to incorporate the effect of the spin fluctuations in the case of an extended interaction $V_{\mathbf{i}\mathbf{j}}$.

The Hubbard-Stratonovich transformation of $n_\mathbf{i}n_\mathbf{j}$ with only density $n_\mathbf{i}$ is exact, but does not take into account electron correlations at the mean-field or at the saddle-point approximation level. Substituting the expression for the density $\hat{n}_\mathbf{i}= \hat{n}_{\mathbf{i},\uparrow} + \hat{n}_{\mathbf{i},\downarrow}$ yields $\hat{n}_\mathbf{i} \hat{n}_\mathbf{j} = \hat{n}_{\mathbf{i},\uparrow}\hat{n}_{\mathbf{j},\uparrow} + \hat{n}_{\mathbf{i},\downarrow}\hat{n}_{\mathbf{j},\downarrow}+ \hat{n}_{\mathbf{i},\uparrow}\hat{n}_{\mathbf{j},\downarrow} + \hat{n}_{\mathbf{i},\downarrow}\hat{n}_{\mathbf{j},\uparrow}$. Choosing the spin direction along the $z$-axis allows one to rewrite the last two terms as $\hat{n}_{\mathbf{i},\uparrow}\hat{n}_{\mathbf{j},\downarrow} + \hat{n}_{\mathbf{i},\downarrow}\hat{n}_{\mathbf{j},\uparrow} =(\hat{n}_\mathbf{i} \hat{n}_\mathbf{j}-4\hat{S}_{\mathbf{i},z}\hat{S}_{\mathbf{j},z})/2$. 
Therefore, 
\begin{equation}
V_{\mathbf{i}\mathbf{j}}\hat{n}_\mathbf{i}\hat{n}_\mathbf{j}=V_{\mathbf{i},\mathbf{j}}(\hat{n}_{\mathbf{i},\uparrow}\hat{n}_{\mathbf{j},\uparrow} + \hat{n}_{\mathbf{i},\downarrow}\hat{n}_{\mathbf{j},\downarrow})+\frac{V_{\mathbf{i}\mathbf{j}}}{2}(\hat{n}_\mathbf{i} \hat{n}_\mathbf{j}-4\hat{S}_{\mathbf{i},z}\hat{S}_{\mathbf{j},z}).
\end{equation}
The first two terms in the parenthesis become trivial when $i=j$, i.e. $\hat{n}_{\mathbf{i},\uparrow}^2=\hat{n}_{\mathbf{i},\uparrow}$ and $\hat{n}_{\mathbf{i},\downarrow}^2=\hat{n}_{\mathbf{i},\downarrow}$. Defining the diagonal component of $V_{\mathbf{i}\mathbf{j}}$ as $U_{\mathbf{i}\mathbf{j}}= U\delta_{\mathbf{i}\mathbf{j}}$, where $U=V_{\mathbf{i}\mathbf{i}}$, we obtain 
\begin{equation}\label{H_int_spin}
 	\hat{H}_{\rm
 		int}=\frac{1}{4}\sum_{\mathbf{i},\mathbf{j}}(2V_{\mathbf{i}\mathbf{j}}-U_{\mathbf{i}\mathbf{j}})\hat{n}_\mathbf{i}\hat{n}_\mathbf{j}-\sum_{\mathbf{i},\mathbf{j}}U_{\mathbf{i}\mathbf{j}}\hat{S}_{\mathbf{i},z}\hat{S}_{\mathbf{j},z}+\frac{1}{2}\sum_{\mathbf{i}}U\hat{n}_{\mathbf{i}},
\end{equation}
where the summation indices run over all lattice sites of all sublattices, i.e.
\begin{equation}
\sum_{\mathbf{i},\mathbf{j}}V_{\mathbf{i}\mathbf{j}}\hat{n}_\mathbf{i}\hat{n}_\mathbf{j}=\sum_{\mathbf{i}\in A,\mathbf{j}\in A}V_{\mathbf{i}\mathbf{j}}\hat{n}_\mathbf{i}\hat{n}_\mathbf{j}+\sum_{\mathbf{i}\in B,\mathbf{j}\in B}V_{\mathbf{i}\mathbf{j}}\hat{n}_\mathbf{i}\hat{n}_\mathbf{j}+2\sum_{\mathbf{i}\in A,\mathbf{j}\in B}V_{\mathbf{i}\mathbf{j}}\hat{n}_\mathbf{i}\hat{n}_\mathbf{j}.
\end{equation}
Assuming translation invariance and defining the Fourier transformations for some $\mathbf{i}\in A$
\begin{equation}
	V^{aa}(\mathbf{k})=\sum_{\mathbf{j}\in A}V_{\mathbf{i}\mathbf{j}}e^{i\mathbf{k} \cdot (\mathbf{j}-\mathbf{i})},
\end{equation}	
and
\begin{equation}
	V^{ab}(\mathbf{k})=\sum_{\mathbf{j}\in B}V_{\mathbf{i}\mathbf{j}}e^{i\mathbf{k} \cdot (\mathbf{j}-\mathbf{i})},
\end{equation}	
with analogous expressions for $V^{bb}(\mathbf{k})$ and $V^{ba}(\mathbf{k})$, with  $V^{ba}(\mathbf{k})=V^{ab}(\mathbf{k})$, the generalized expression for $\mathbf{U}$ becomes
\begin{equation}
	\mathbf{U}_{(\omega_n,\mathbf{k}),(\omega_{n'},\mathbf{k}')}=\frac{	\delta_{n,-n'}\delta_{\mathbf{k},-\mathbf{k}'}}{2}\left(\begin{matrix}
		U-2 V^{aa}(\mathbf{k})& 0 & -2 V^{ab}(\mathbf{k}) & 0 \\
		0 & U\mathbf{m} & 0 & 0 \\
		-2 V^{ba}(\mathbf{k}) & 0 & U-2 V^{bb}(\mathbf{k})& 0 \\
		0 & 0 & 0 & U\mathbf{m}
	\end{matrix}\right).
\end{equation}
Absorbing $U/2$ from the last term of Eq.\;(\ref{H_int_spin}) in the redefinition of the chemical potential $\mu$, the partition function is again expressed in the generic form given by Eq.\;(\ref{Z_psi_M}). 
\subsection{Correlation functions}
The two-point correlation function is generated by the variation $\delta\mathbf{h}$ of the Green's function parameter $\mathbf{h}$,  
\begin{equation}
Z^{-1}[\mathbf{h}] \delta Z[\mathbf{h}]=Z^{-1}[\mathbf{h}]\int \mathrm{d}[\psi^{\dag}]\mathrm{d}[\psi]\mathrm{d}[\mathbf{M}]\psi^{\dag}\delta\mathbf{h}\psi\exp\left(\psi^{\dag}\mathbf{G}^{-1}(\mathbf{M})\psi-\frac{N\beta}{ 2 }\mathbf{U}^{-1}\mathbf{M}\mathbf{M}\right),
\end{equation}
taking into account that 
\begin{equation}\nonumber
\delta \mathbf{G}^{-1}(\mathbf{M})=\delta\mathbf{h}
\end{equation}
and evaluating at $\mathbf{h}=0$.
The correlation function is directly connected to the variation of $\ln(Z[\mathbf{h}])$, namely, $Z^{-1}[\mathbf{h}]\delta Z[\mathbf{h}]=\delta  \ln(Z[\mathbf{h}])$.
The four-point correlation function is generated by the two variations $\delta\mathbf{h}_1$ and $\delta\mathbf{h}_2$, 
\begin{equation}
Z^{-1}[\mathbf{h}] \delta^2_{\mathbf{h}_1\mathbf{h}_2} Z[\mathbf{h}]=Z^{-1}[\mathbf{h}] \int \mathrm{d}[\psi^{\dag}]\mathrm{d}[\psi]\mathrm{d}[\mathbf{M}](\psi^{\dag}\delta_{\mathbf{h}_1}\psi)(\psi^{\dag}\delta_{\mathbf{h}_2}\psi)\exp\left(\psi^{\dag}\mathbf{G}^{-1}(\mathbf{M})\psi-\frac{N\beta}{ 2 }\mathbf{U}^{-1}\mathbf{M}\mathbf{M}\right),
\end{equation}
which is 
\begin{equation}\label{secondderiv}
Z^{-1}[\mathbf{h}] \delta^2_{\mathbf{h}_1\mathbf{h}_2} Z[\mathbf{h}]=\delta^2_{\mathbf{h}_1\mathbf{h}_2}
 \ln(Z[\mathbf{h}])+\delta_{\mathbf{h}_1}  \ln(Z[\mathbf{h}])\; \delta_{\mathbf{h}_2}  \ln(Z[\mathbf{h}]).
\end{equation}
$Z[\mathbf{h}]$ can be expressed as a partition function for only bosonic fields $\mathbf{M}$ after integrating out the fermionic fields. The integration  yields
\begin{equation}\label{fermionic_det}
\int \mathrm{d}[\psi^{\dag}]\mathrm{d}[\psi]\exp(\psi^{\dag}\mathbf{G}^{-1}(\mathbf{M})\psi) = \det[-\mathbf{G}^{-1}(\mathbf{M})].
\end{equation}
Substituting Eq.\;(\ref{fermionic_det}) into Eq.\;(\ref{Z_psi_M}) leads to 
\begin{equation}
Z[\mathbf{h}]=\int \mathrm{d}[\mathbf{M}]\exp\left(-\frac{N\beta}{ 2 }\mathbf{U}^{-1}\mathbf{M}
\mathbf{M}+{\rm Tr}\{\ln[-\mathbf{G}^{-1}(\mathbf{M})]\}\right).
\end{equation}
\end{widetext}
In the noninteracting case, the partition function is given by
\begin{equation}
\ln(Z[\mathbf{h}])={\rm Tr}[\ln(-\mathbf{G}^{-1}(0))],
\end{equation}
and the variation equation simplifies to just
\begin{equation}
\delta\ln(Z[\mathbf{h}])={\rm
Tr}[\mathbf{G}(0)\delta\mathbf{h}]
\end{equation}
and
\begin{equation}
\delta^2 \ln(Z[\mathbf{h}])=-{\rm
Tr}[\mathbf{G}(0)\delta\mathbf{h}_2\mathbf{G}(0)\delta\mathbf{h}_1],
\end{equation}
leading to the well known Wick's formula
\begin{eqnarray}\label{Wicksformula}
Z^{-1}[\mathbf{h}] \delta^2 Z[\mathbf{h}]&=&{\rm
Tr}[\mathbf{G}(0)\delta\mathbf{h}_2]{\rm
Tr}[\mathbf{G}(0)\delta\mathbf{h}_1]  \\ 
&-& {\rm Tr}[\mathbf{G}(0)\delta\mathbf{h}_2\mathbf{G}(0)\delta\mathbf{h}_1]. \nonumber
\end{eqnarray}

\subsection{Free energy and correlation functions}

We denote $Z(f)=Z[\mathbf{h}(f)]$, where $\mathbf{h}(f) = f \mathbf{Q}$. Here, $\mathbf{Q}$ defines the two-point correlation function describing the coupling between fermions. The corresponding order parameter is given by
\begin{equation}
\phi_f=\frac{\mathrm{d}}{\mathrm{d}
f}\ln[Z(f)],
\end{equation}
which we can consider as a conjugate variable to $f$, corresponding to the expectation value 
\begin{equation}\label{order_param}
\phi_f=\langle \psi^{\dag}\mathbf{Q}\psi
\rangle_f.
\end{equation}
If $\mathbf{Q}$ is the identity matrix, Eq.\;(\ref{order_param}) becomes the expected density; if $\mathbf{Q}$ is proportional to a Pauli matrix, Eq.\;(\ref{order_param}) leads to the expected spin density. The corresponding expectation value and higher-order correlation functions can be found from the generating functional
\begin{eqnarray} \nonumber 
\phi_f &=& 
Z^{-1}\int \mathrm{d}[\psi^{\dag}]\mathrm{d}[\psi]\mathrm{d}[\mathbf{M}]\exp\left(\psi^{\dag}\mathbf{G}^{-1}(\mathbf{M})\psi  \right.\\
&-&  \left. \frac{N\beta}{ 2 }\mathbf{U}^{-1}\mathbf{M}\mathbf{M}\right)\psi^{\dag}\mathbf{Q}\psi.
\end{eqnarray}
The information about a phase transition of the system is implicitly encoded in the partition function. Thus, to make it evident, we have
to find a free energy depending on the order parameter by performing
the Legendre transformation
\begin{equation}
\beta F(\phi_f)=\phi_f f -\ln[Z(f)],
\end{equation}
which also implies that
\begin{equation}\nonumber
f =\frac{\mathrm{d}}{\mathrm{d} \phi_f}\beta F(\phi_f).
\end{equation}
For small perturbations, it holds that $\phi_f = \phi_0 + \alpha^{-1} f +\ldots$, where the inverse of $\alpha$ is the susceptibility, given by
\begin{equation}
\alpha^{-1}=\frac{\mathrm{d}^2}{{\mathrm{d} f}^2}\ln(Z(f))
|_{f=0}.
\end{equation}
Consequently,  the free energy can be expressed as
\begin{equation}
\beta F[\phi_f]=\beta F[\phi_0]+\frac{1}{2}\alpha(\phi_f-\phi_0)^2+\ldots.
\end{equation}
Clearly, if $\alpha$ becomes negative, the system becomes unstable and undergoes a
second-order phase transition. The instability condition is
$\alpha=0$. The susceptibility is related to the correlation functions as follows
\begin{equation}
\alpha^{-1}=\langle(\psi^{\dag}\mathbf{Q}\psi)^2\rangle-\langle \psi^{\dag}\mathbf{Q}\psi\rangle^2.
\end{equation}
The instability implies that
$\alpha^{-1}$ diverges and so does the correlation function $\langle
(\psi^{\dag}\mathbf{Q}\psi)^2\rangle$, signalling the second-order phase transition. In the noninteracting case,
the expressions become $\phi_f={\rm	Tr}[\mathbf{G}\mathbf{Q}]$ and $\alpha^{-1}=-{\rm	Tr}[\mathbf{G}\mathbf{Q}\mathbf{G}\mathbf{Q}]$.
At this point, we notice that knowing the partition function $Z[\mathbf{h}]$ for an arbitrary matrix $\mathbf{h}$ is completely sufficient to find all the relevant correlation functions, including those for Cooper pairs, as shown below,
\begin{eqnarray} && \nonumber
\int_0^{\hbar\beta} \ud \tau \int_0^{\hbar\beta} \ud \tau'
a^{\dag}_{\mathbf{k},\uparrow}(\tau)a^{\dag}_{-\mathbf{k},\downarrow}(\tau)
a_{-\mathbf{k}',\downarrow}(\tau')a_{\mathbf{k}',\uparrow}(\tau') \\
&=& \sum_{\omega_{n},\omega_{n'}}a^{\dag}_{\omega_{n},\mathbf{k},\uparrow}
a^{\dag}_{-\omega_{n},-\mathbf{k},\downarrow}a_{-\omega_{n'},-\mathbf{k}',\downarrow}a_{\omega_{n'},\mathbf{k}',\uparrow},
\end{eqnarray}
namely,
\begin{equation}\label{cooper_corrfunc}
\sum_{\mathbf{q}_1,\mathbf{q}_2} \left\langle a^{\dag}_{\mathbf{q}_1,\uparrow}a^{\dag}_{-\mathbf{q}_1,\downarrow}a_{\mathbf{q}_2,\uparrow}a_{-\mathbf{q}_2,\downarrow}\right\rangle.
\end{equation}
The correlation function is generated when adding a source term of the form 
\begin{equation}\label{cooper_source}
\sum_{\mathbf{q}_1,\mathbf{q}_2} a^{\dag}_{\mathbf{q}_1,\uparrow}\Delta^{*}_{\mathbf{q}_1,-\mathbf{q}_1}a^{\dag}_{-\mathbf{q}_1,\downarrow}a_{\mathbf{q}_2,\uparrow}\Delta_{\mathbf{q}_2,-\mathbf{q}_2}a_{-\mathbf{q}_2,\downarrow}
\end{equation}
and differentiating with respect to $\Delta$ and $\Delta^{*}$. Equivalently, rearranging the terms, 
\begin{equation}
	\sum_{\mathbf{q}_1,\mathbf{q}_2} a^{\dag}_{\mathbf{q}_1,\uparrow}a_{-\mathbf{q}_2,\downarrow}a^{\dag}_{-\mathbf{q}_1,\downarrow}a_{\mathbf{q}_2,\uparrow}\Delta^{*}_{\mathbf{q}_1,-\mathbf{q}_1}\Delta_{\mathbf{q}_2,-\mathbf{q}_2},
\end{equation}
this correlation function is generated by varying  
the partition function with the following source terms by $\mathbf{h}_{+}$ and $\mathbf{h}_{-}$,
\begin{equation}
\sum_{\mathbf{q}_1,\mathbf{q}_2}(\psi^{\dag}_{\mathbf{q}_1}\mathbf{P}^{a}_{+}\psi_{-\mathbf{q}_2}h^{a}_{\mathbf{q}_1,-\mathbf{q}_2,+}+\psi^{\dag}_{-\mathbf{q}_1}\mathbf{P}^{a}_{-}\psi_{\mathbf{q}_2}h^{a}_{-\mathbf{q}_1,\mathbf{q}_2,-}),
\end{equation}
where $\mathbf{P}^{a}_{+}$ and $\mathbf{P}^{a}_{-}$ are such that $\psi^{\dag}_{\mathbf{q}_1}\mathbf{P}^{a}_{+}\psi_{-\mathbf{q}_2}= a^{\dag}_{\mathbf{q}_1,\uparrow}a_{-\mathbf{q}_2,\downarrow}$ and $\psi^{\dag}_{-\mathbf{q}_1}\mathbf{P}^{a}_{-}\psi_{\mathbf{q}_2}= a^{\dag}_{-\mathbf{q}_1,\downarrow}a_{\mathbf{q}_2,\uparrow}$.
This allows us to investigate the stability of the system with respect to the Cooper instability, which is signalled by a divergence of the correlation function.

\section{Partition function evaluation}
In order to evaluate the partition function integral,
\begin{equation}
Z[\mathbf{h}]=\int \mathrm{d}[\mathbf{M}]\exp(-{\cal S}(\mathbf{M})/\hbar),
\end{equation}
where
\begin{equation}\label{HS_action}\nonumber
{\cal S}(\mathbf{M})=\frac{N\hbar\beta}{ 2 }\mathbf{U}^{-1}\mathbf{M}
\mathbf{M}-\hbar{\rm Tr}[\ln(-\mathbf{G}^{-1}(\mathbf{M})),
\end{equation}
we are going to approximate the integrand as an exponential of a quadratic function, such that the problem is reduced to a Gaussian integral. 

\subsection{Saddle-point approximation}
Following the saddle point approximation, the quadratic function is chosen such that it matches ${\cal S}(\mathbf{M})$ in the vicinity of a saddle point $\mathbf{M}_{\rm sp}$. For a quadratic function of the form  
\begin{equation}
{\cal S}_{\rm sp}(\mathbf{M})={\cal S}_{\rm
sp}+\frac{N\hbar\beta}{ 2 }\mathbf{W}^{-1} \mathbf{Y}(\mathbf{M}) \mathbf{Y}(\mathbf{M}),
\end{equation}
with $\mathbf{Y} = \mathbf{M} - \mathbf{M}_{\rm sp}$,
the matching constraint for the difference 
\begin{equation}\nonumber
N\hbar\beta L(\mathbf{M}) \equiv {\cal S}(\mathbf{M})-{\cal S}_{\rm sp}(\mathbf{M})
\end{equation}
reads as $L(\mathbf{M}_{\rm sp})=0$, $\mathrm{d}L(\mathbf{M}_{\rm sp})=0$, and $\mathrm{d}^2 L(\mathbf{M}_{\rm sp})=0$. This implies that
\begin{equation}\nonumber
{\cal S}_{\rm sp}={\cal S}(\mathbf{M}_{\rm sp}),
\end{equation}
whereas the first differential is given by
\begin{equation}\nonumber
\ud L(\mathbf{M})=\mathbf{U}^{-1}\mathbf{M}\ud\mathbf{M} -\frac{1}{ N\beta }{\rm
Tr}[\mathbf{G}\ud\mathbf{G}^{-1}]-\mathbf{W}^{-1} \mathbf{Y}
\ud\mathbf{Y}
\end{equation}
and the second differential is
\begin{equation}\nonumber
\ud^2 L(\mathbf{M})=\mathbf{U}^{-1}\ud \mathbf{M}\ud \mathbf{M}+\frac{1}{ N\beta}{\rm
Tr}[(\mathbf{G}\ud\mathbf{G}^{-1})^2]-\mathbf{W}^{-1} \ud\mathbf{Y}\ud\mathbf{Y},
\end{equation}
where we utilize the following observation
\begin{equation}\nonumber
{\rm Tr}[\ud^n\ln(-\mathbf{G}^{-1})]=(-1)^{n-1}(n-1)!{\rm
Tr}[(\mathbf{G}\ud\mathbf{G}^{-1})^n].
\end{equation}
Evaluating these expressions at the saddle point $\mathbf{M} = \mathbf{M}_{\rm sp}$ and taking into account that $\mathbf{Y}(\mathbf{M}_{\rm sp})=0$ and $\ud\mathbf{Y} = \ud\mathbf{M}$, we obtain
\begin{equation}\label{HF}
\mathbf{U}^{-1}\mathbf{M}_{\rm sp} =\Gamma^{(1)}
\end{equation}
and
\begin{equation}\label{RPA}
\mathbf{W}^{-1} =\mathbf{U}^{-1}-\Gamma^{(2)}.
\end{equation}
For convenience, we introduce covariant tensors $\Gamma^{(n)}$ that are totally symmetric in the pairs of indices $(\mathbf{p},r)$, $(\mathbf{p}',r')$, $\ldots$, $(\mathbf{p}^n,r^n)$,  to denote
\begin{equation}\label{Gamma_def}
\frac{(-1)^{n-1}(n-1)!}{ \beta N }{\rm
Tr}[(\mathbf{G}(\mathbf{M}_{\rm sp})\ud\mathbf{G}^{-1})^n]=\Gamma^{(n)}\overbrace{\ud\mathbf{M}\ldots\ud\mathbf{M}}^{n}.
\end{equation}
The explicit expressions are
\begin{align}
	\begin{split}
		\Gamma^{(1)}_{(\mathbf{p},r)} &= \frac{1}{ \beta N }\sum_{\mathbf{q}}{\rm
			Tr}[\mathbf{GR}_{\mathbf{q},\mathbf{q};\mathbf{p},r}] \\ \nonumber
		\Gamma^{(2)}_{(\mathbf{p},r),(\mathbf{p}',r')}& = \frac{-1}{ \beta N }\sum_{\mathbf{q},\mathbf{q}'}{\rm
			Tr}[\mathbf{GR}_{\mathbf{q},\mathbf{q'};\mathbf{p},r}\mathbf{GR}_{\mathbf{q}',\mathbf{q};\mathbf{p}',r'}], \\\nonumber
	\end{split}
\end{align}
where we denote the product
\begin{equation}\nonumber
\mathbf{GR}_{\mathbf{q},\mathbf{q}';\mathbf{p},r} = \sum_{\mathbf{q}''}\mathbf{G}_{\mathbf{q},\mathbf{q}''}\mathbf{R}_{\mathbf{q}'',\mathbf{q}';\mathbf{p},r}=\mathbf{G}_{\mathbf{q},\mathbf{q}'+\mathbf{p}}\mathbf{R}_{\mathbf{q}'+\mathbf{p},\mathbf{q}';\mathbf{p},r},
\end{equation}
and, analogously,
\begin{equation}\label{RG_def}
\mathbf{RG}_{\mathbf{q},\mathbf{q}';\mathbf{p},r} = \sum_{\mathbf{q}''}\mathbf{R}_{\mathbf{q},\mathbf{q}'';\mathbf{p},r}\mathbf{G}_{\mathbf{q}'',\mathbf{q}'}=\mathbf{R}_{\mathbf{q},\mathbf{q}-\mathbf{p};\mathbf{p},r}\mathbf{G}_{\mathbf{q}-\mathbf{p},\mathbf{q}'}.
\end{equation}
Here, Eq.\;(\ref{HF}) can be recognized as a Hartree-Fock approximation, i.e. $\mathbf{M}_{\rm sp} =\mathbf{U}\Gamma^{(1)}$ and Eq.\;(\ref{RPA}) as an RPA result for the effective interaction. Now, we recall that we defined the Hubbard-Stratonovich transformation with Gaussian integrals having a certain offset value in the integration contours, which we denote $\mathbf{M}_{\rm sp}$. The freedom of choosing the value of $\mathbf{M}_{\rm sp}$, which can be also complex, comes from the well known invariance of the Gaussian integrals with respect to a constant shift of integration variables. We will specify the value of $\mathbf{M}_{\rm sp}$ later, but in most cases it turns out to be real in coordinate representation. Assuming the value is real in
coordinate space, its Fourier transform satisfies
$\mathbf{M}_{{\rm sp}}^{-\mathbf{p}}=(\mathbf{M}_{{\rm sp}}^{\mathbf{p}})^{*}$.
However, this relation holds only for the offset value  $\mathbf{M}_{\rm sp}$, since the integration for the components of the Hubbard-Stratonovich fields coupled with density goes along a
contour parallel to the imaginary axis from $-i\infty$ to
$+i\infty$. This implies that $\mathbf{Y}^{-\mathbf{p}}=\eta(\mathbf{Y}^{\mathbf{p}})^{*}$,
i.e. $\mathbf{Y}=\tilde{\eta}\mathbf{Y}^{*}$, where the symmetric covariant tensor
$\tilde{\eta}_{\mathbf{p},\mathbf{p}'}=\eta
\delta_{\mathbf{p},-\mathbf{p}'}$. This property implies
\begin{equation}\nonumber
Z_{\rm sp}[\mathbf{h}]=\int \mathrm{d}[\mathbf{Y}]\exp\left(-\frac{1}{\hbar}{\cal S}(\mathbf{M}_{\rm sp})-\frac{N\beta}{ 2 }\mathbf{W}^{-1}\mathbf{Y}
(\tilde{\eta}\mathbf{Y}^{*})\right).
\end{equation}
Clearly, in this case we can perform Gaussian integration to find
the partition function
\begin{equation}
Z_{\rm sp}[\mathbf{h}]=\exp(-{\cal S}(\mathbf{M}_{\rm sp})/\hbar)\det[\mathbf{U}\mathbf{W}^{-1}]^{-1/2},
\end{equation}
or more explicitly,
\begin{eqnarray}
Z_{\rm sp}[\mathbf{h}]&=&\exp\left(-\frac{N\beta}{ 2
}\mathbf{U}^{-1}\mathbf{M}_{\rm sp}\mathbf{M}_{\rm sp} \right. \\  \nonumber
&+& \left. {\rm
Tr}[\ln(-\mathbf{G}^{-1}(\mathbf{M}_{\rm sp}))]-\frac{1}{2}{\rm Tr}[\ln(\mathbf{U}\mathbf{W}^{-1})]\right),
\end{eqnarray}
which in combination with the expressions for $\mathbf{M}_{\rm sp}$ and $\mathbf{W}^{-1}$ determine $Z_{\rm sp}[\mathbf{h}]$. In the following,  we denote $Z_{\rm sp}[\mathbf{h}]$ by $Z$ for simplicity.  As shown above, it is beneficial to operate with $\ln(Z)$, which is
\begin{eqnarray} \nonumber 
\ln(Z)&=&-\frac{N\beta}{ 2
}\mathbf{U}^{-1}\mathbf{M}_{\rm sp}\mathbf{M}_{\rm sp}+{\rm
Tr}[\ln(-\mathbf{G}^{-1}(\mathbf{M}_{\rm sp}))] \\
&-& \frac{1}{2}{\rm Tr}[\ln(\mathbf{U}\mathbf{W}^{-1})].
\end{eqnarray}
Now, we observe that
\begin{equation}
\delta\mathbf{G}^{-1}(\mathbf{M}_{\rm sp})= \mathbf{R}\delta\mathbf{M}_{\rm sp}+ \delta\mathbf{h},
\end{equation}
which leads to
\begin{equation}\label{delta_gamma}
\delta  \Gamma^{(n)}= \delta_{\mathbf{h}}  \Gamma^{(n)}+ \Gamma^{(n+1)}\delta\mathbf{M}_{\rm sp},
\end{equation}
where $\delta_{\mathbf{h}}  \Gamma^{(n)}$ denotes variation only with respect to $\mathbf{h}$, keeping $\mathbf{M}_{\rm sp}$ fixed, i.e. $\delta_{\mathbf{h}} \mathbf{G}^{-1}=\delta\mathbf{h}$. This can be calculated by using the identity $\delta\mathbf{G} = -\mathbf{G}\delta\mathbf{G}^{-1}\mathbf{G}$. In particular,
\begin{equation}\label{delta_log_det}
\delta_{\mathbf{h}}{\rm Tr}[\ln(-\mathbf{G}^{-1}(\mathbf{M}_{\rm sp}))]={\rm Tr}[\mathbf{G}(\mathbf{M}_{\rm sp})\delta\mathbf{h}].
\end{equation}
Using Eq.\;(\ref{delta_gamma}) and Eq.\;(\ref{delta_log_det}), we find that the first variation of  $\ln(Z)$ with respect to $\delta\mathbf{h}$ reads
\begin{eqnarray} \nonumber 
\delta  \ln(Z[\mathbf{h}])&=&-N\beta\mathbf{U}^{-1}\mathbf{M}_{\rm sp}\delta \mathbf{M}_{\rm sp}+N\beta\Gamma^{(1)}\delta \mathbf{M}_{\rm sp} \\
&+&{\rm Tr}[\mathbf{G}(\mathbf{M}_{\rm sp})\delta\mathbf{h}]-\frac{1}{2}{\rm Tr}[\mathbf{W} \delta\mathbf{W}^{-1}],
\end{eqnarray}
and can be simplified by using $\mathbf{U}^{-1}\mathbf{M}_{\rm sp}\delta \mathbf{M}_{\rm sp} =\Gamma^{(1)}\delta \mathbf{M}_{\rm sp}$ to
\begin{equation}
\delta \ln(Z[\mathbf{h}])={\rm
Tr}[\mathbf{G}(\mathbf{M}_{\rm sp})\delta\mathbf{h}]-\frac{1}{2}{\rm Tr}[\mathbf{W} \delta\mathbf{W}^{-1}].
\end{equation}
This closed form solution directly allows for variation calculation
\begin{equation}\label{delta_msp}
\mathbf{U}^{-1}\delta\mathbf{M}_{\rm sp} =\delta_{\mathbf{h}}\Gamma^{(1)}+\Gamma^{(2)}\delta\mathbf{M}_{\rm sp} ,
\end{equation}
and
\begin{equation}\label{delta_w}
\delta\mathbf{W}^{-1} =-\delta_{\mathbf{h}}\Gamma^{(2)}-\Gamma^{(3)}\delta\mathbf{M}_{\rm sp},
\end{equation}
where we used Eq.\;(\ref{delta_gamma}) to obtain Eqs.\;(\ref{delta_msp}) and (\ref{delta_w}). 
Eq.\;(\ref{delta_msp}) directly yields
\begin{equation}
\delta\mathbf{M}_{\rm sp} =\mathbf{W}\delta_{\mathbf{h}}\Gamma^{(1)},
\end{equation}
which after substitution into Eqs.\;(\ref{delta_w}) provides
\begin{equation}
\delta\mathbf{W}^{-1} =-\delta_{\mathbf{h}}\Gamma^{(2)}-\Gamma^{(3)}\mathbf{W}\delta_{\mathbf{h}}\Gamma^{(1)}.
\end{equation}
Thus, the first variation reads as
\begin{eqnarray} \nonumber 
\delta \ln(Z[\mathbf{h}]) &=& {\rm
Tr}[\mathbf{G}(\mathbf{M}_{\rm sp})\delta\mathbf{h}]+\frac{1}{2}{\rm Tr}[\mathbf{W} \delta_{\mathbf{h}}\Gamma^{(2)}] \\ &+& \frac{1}{2}{\rm Tr}[\mathbf{W} \Gamma^{(3)}\mathbf{W}\delta_{\mathbf{h}}\Gamma^{(1)}].
\end{eqnarray}
In particular, for $\mathbf{h}(f)=f\mathbf{Q}$, the first term becomes
${\rm	Tr}[\mathbf{G}(\mathbf{M}_{\rm sp})\delta\mathbf{h}] = {\rm
	Tr}[\mathbf{G}\mathbf{Q}]\delta f$.
The second variation leads to
\begin{equation}\nonumber
\delta {\rm
Tr}[\mathbf{G}(\mathbf{M}_{\rm sp})\delta\mathbf{h}]=-{\rm
Tr}[\mathbf{G}\mathbf{Q}\mathbf{G}\mathbf{Q}] ({\delta f})^2 -N\beta\mathbf{W}^{-1}\delta\mathbf{M}_{\rm sp}\delta\mathbf{M}_{\rm sp}
\end{equation}
or, equivalently
\begin{equation}\nonumber
\delta {\rm
Tr}[\mathbf{G}(\mathbf{M}_{\rm sp})\delta\mathbf{h}]=\beta N\kappa ({\delta f})^2 -N\beta\mathbf{W}\delta_{\mathbf{h}}\Gamma^{(1)}\delta_{\mathbf{h}}\Gamma^{(1)},
\end{equation}
where 
\begin{equation}\nonumber
\delta_{\mathbf{h}}\Gamma^{(1)} = \frac{-\delta f}{ \beta N }{\rm
	Tr}[\mathbf{GR}\mathbf{G}\mathbf{Q}],
\end{equation}
and we denote
\begin{equation}\label{kappa}
\kappa=\frac{-1}{ \beta N }{\rm
Tr}[\mathbf{G}\mathbf{Q}\mathbf{G}\mathbf{Q}].
\end{equation}

Clearly, within the saddle-point approximation, the second variation and the susceptibility remain finite for non-singular $\mathbf{W}$. In order to investigate an instability with non-singular $\mathbf{W}$, a more advanced approximation is required. This is described in the next section.  

\begin{widetext}
\subsection{Median-point approximation}
As a next step, we apply a generalized saddle-point approximation to evaluate the partition function. The main idea of the method is to perform  a change in the integration variables before applying the saddle-point approximation, to make it more efficient. The concept is illustrated for the simplest case of one-dimensional integrals in Appendix \ref{one_dimensional_integrals}. 
Going back to the partition function, we follow the same approach as in the previous section and consider 
the Gaussian integral
\begin{equation}
Z_{\rm sp}=\int \mathrm{d}[\mathbf{Y}]\exp\left(-\frac{{\cal S}_{\rm
sp}}{\hbar}-\frac{N\beta}{ 2 }\mathbf{W}^{-1} \mathbf{Y} \mathbf{Y}\right),
\end{equation}
where $\mathbf{W}^{-1}$ is some fixed covariant tensor that does not depend on $\mathbf{Y}$ and the variables $\mathbf{Y}$ satisfy
$\mathbf{Y}^{-\mathbf{p}}=\eta(\mathbf{Y}^{\mathbf{p}})^{*}$. 
Furthermore, we consider a change of variables $\mathbf{Y}\rightarrow\mathbf{M}$ in the integral, with
$\mathrm{d}\mathbf{Y}(\mathbf{M})=\mathbf{Y}'(\mathbf{M})\mathrm{d}\mathbf{M}$.
The notation implies
\begin{equation}
\label{differantialindex} \ud Y^{( \mathbf{p},r)
}=Y'^{( \mathbf{p},r)}_{(\mathbf{p}',r')}\ud
M^{(\mathbf{p}',r')}.
\end{equation}

The change of variables leads to
\begin{equation}
Z_{\rm sp}=\int \mathrm{d}[\mathbf{M}]\det[\mathbf{Y}'(\mathbf{M})]\exp\left[-\frac{{\cal S}_{\rm
sp}}{\hbar}-\frac{N\beta}{ 2 }\mathbf{W}^{-1} \mathbf{Y}(\mathbf{M}) \mathbf{Y}(\mathbf{M})\right].
\end{equation}
Therefore, the power of the exponent transforms from the quadratic function into  
\begin{equation}\label{SP_action}
{\cal S}_{\rm sp}(\mathbf{M})={\cal S}_{\rm
sp}+\frac{N\hbar\beta}{ 2 }\mathbf{W}^{-1} \mathbf{Y}(\mathbf{M}) \mathbf{Y}(\mathbf{M}))-\hbar{\rm
Tr}[\ln(\mathbf{Y}')],
\end{equation}
where the last term comes from the Jacobian of the transformation. Choosing for convenience the integration variables such that  $\mathbf{Y}'(\mathbf{M}_{\rm sp})=\mathbf{I}$ at the saddle point $\mathbf{Y}(\mathbf{M}_{\rm sp})=0$, we can perform the Gaussian integration to find
\begin{equation}
Z_{\rm sp}=\exp(-{\cal S}_{\rm sp}/\hbar)\det[\mathbf{U}\mathbf{W}^{-1}]^{-1/2}.
\end{equation} 

Usually, the standard saddle-point approximation prescribes to first find the saddle point, and then to perform a quadratic expansion around it. Here we adopt a different procedure, namely we consider integrals (partition functions) that can be exactly mapped into the Gaussian integral by performing the appropriate change of variables.
  Analogously to the saddle-point approximation, the function ${\cal S}_{\rm sp}(\mathbf{M})$ in Eq.\;(\ref{SP_action}) is chosen such that it matches ${\cal S}(\mathbf{M})$ in Eq.\;(\ref{HS_action}) in the vicinity of the saddle point $\mathbf{Y}(\mathbf{M}_{\rm sp})=0$ or, equivalently, the matching constraint for the difference 
\begin{equation}\label{action_constraint}
N\hbar\beta L(\mathbf{M}) \equiv {\cal S}(\mathbf{M})-{\cal S}_{\rm sp}(\mathbf{M})
\end{equation}
vanishes in the vicinity of $\mathbf{M}_{\rm sp}$, namely, $L(\mathbf{M}_{\rm sp})=0$, $\mathrm{d}L(\mathbf{M}_{\rm sp})=0$, and all the higher differentials up to a certain order $k$, $\mathrm{d}^k L(\mathbf{M}_{\rm sp})=0$. This implies that
\begin{equation}\label{SP_action_const}
{\cal S}_{\rm sp}={\cal S}(\mathbf{M}_{\rm sp}).
\end{equation}
Therefore, in 1D $\mathbf{M}_{\rm sp}$ would be the median point of the distribution, which coincides with the saddle point in the Gaussian case. 
Substitution of the value ${\cal S}_{\rm sp}$ results into
\begin{equation}
Z=\exp\left\{\frac{-N\beta}{ 2}\mathbf{U}^{-1}\mathbf{M}_{\rm sp}\mathbf{M}_{\rm sp}+{\rm
Tr}[\ln(-\mathbf{G}^{-1}(\mathbf{M}_{\rm sp}))]-\frac{1}{2}{\rm Tr}[\ln(\mathbf{U}\mathbf{W}^{-1})]\right\},
\end{equation}
where we have taken into account that ${\rm Tr}[\ln(\mathbf{Y}'(\mathbf{M}_{\rm sp}))]=0$.
More explicitly, the constraint is obtained by substituting Eq.\;(\ref{HS_action}), Eq.\;(\ref{SP_action}) and Eq.\;(\ref{SP_action_const}) into Eq.\;(\ref{action_constraint}),
\begin{equation}
N\hbar\beta L(\mathbf{M}) = \frac{N\hbar\beta}{ 2 }\mathbf{U}^{-1}\mathbf{M}
\mathbf{M}-\hbar{\rm Tr}[\ln(-\mathbf{G}^{-1})]+\hbar{\rm
Tr}[\ln(\mathbf{Y}')]-{\cal S}(\mathbf{M}_{\rm sp})-\frac{N\hbar\beta}{ 2 }\mathbf{W}^{-1} \mathbf{Y} \mathbf{Y}.
\end{equation}

In general, we denote
\begin{equation}
\label{higherdifferantialindex}
\ud^K Y^{(\mathbf{p},r)}=Y^{(K)\; (\mathbf{p},r)}_{(\mathbf{p}^{1},r^{1})\ldots (\mathbf{p}^{K},r^{K})}\ud M^{(\mathbf{p}^{K},r^{K})}\ldots \ud M^{(\mathbf{p}^{1},r^{1})}
\end{equation}
as
\begin{equation}
\ud^K\mathbf{Y}=\mathbf{Y}^{(K)}\overbrace{\ud\mathbf{M}\ldots\ud\mathbf{M}}^{K}.
\end{equation}
Moreover, for convenience we also denote
\begin{equation}
\ud\mathbf{M}^{K} = \overbrace{\ud\mathbf{M}\ldots\ud\mathbf{M}}^{K}.
\end{equation}

For the more general cases of tensor products, we assume pairing and contraction of nearest indices, i.e. for a covariant tensor $\mathbf{A}$ and a contravariant $\mathbf{B}$
\begin{equation}
\mathbf{A}\mathbf{B}=A_{(\mathbf{p}^{1},r^{1})\ldots (\mathbf{p}^{m},r^{m})}B^{(\mathbf{p}^{m},r^{m})\ldots (\mathbf{p}^{n},r^{n})}
\end{equation}
or
\begin{equation}
\mathbf{B}\mathbf{A}=B^{(\mathbf{p}^{1},r^{1})\ldots (\mathbf{p}^{m},r^{m})}A_{(\mathbf{p}^{m},r^{m})\ldots (\mathbf{p}^{n},r^{n})}.
\end{equation}
The product of a covariant tensor and the $\mathbf{Y}^{(K)}$ derivative  implies
\begin{equation}
\mathbf{A}\mathbf{Y}^{(K)}=A_{(\mathbf{p}^{1},r^{1})\ldots (\mathbf{p}^{m},r^{m})}Y^{(K)\; (\mathbf{p}^{m},r^{m})}_{(\mathbf{q}^{1},r^{1})\ldots (\mathbf{q}^{K},r^{K})}
\end{equation}
and the product of a contravariant tensor and the $\mathbf{Y}^{(K)}$ derivative  implies
\begin{equation}
\mathbf{Y}^{(K)}\mathbf{B}=Y^{(K)\; (\mathbf{q},s)}_{(\mathbf{p}^{1},r^{1})\ldots (\mathbf{p}^{K},r^{K})}B^{(\mathbf{p}^{K},r^{K})\ldots (\mathbf{q}^{n},r^{n})}.
\end{equation}

The first differential of $L(\mathbf{M})$ is given by
\begin{equation}
\ud L(\mathbf{M})=\mathbf{U}^{-1}\mathbf{M}\ud\mathbf{M} -\frac{1}{ N\beta }{\rm
Tr}[\mathbf{G}\ud\mathbf{G}^{-1}]+\frac{1}{ N\beta } {\rm
Tr}[\ud\ln(\mathbf{Y}')]-\mathbf{W}^{-1} \mathbf{Y}
\ud\mathbf{Y},
\end{equation}
with
\begin{equation}
{\rm Tr}[\ud\ln(\mathbf{Y}')]={\rm Tr}[(\mathbf{Y}')^{-1}\ud\mathbf{Y}'].\nonumber
\end{equation}
The second differential reads
\begin{equation}
\ud^2 L(\mathbf{M})=\mathbf{U}^{-1}\ud \mathbf{M}^2+\frac{1}{ N\beta}{\rm
Tr}[(\mathbf{G}\ud\mathbf{G}^{-1})^2]+\frac{1}{ N\beta } {\rm
Tr}[\ud^2\ln(\mathbf{Y}')]-\mathbf{W}^{-1} (\ud\mathbf{Y}^2+\mathbf{Y}\ud^2\mathbf{Y}),
\end{equation}
with
\begin{equation}
{\rm Tr}[\ud^2\ln(\mathbf{Y}')]={\rm Tr}{(\mathbf{Y}')^{-1}\ud^2\mathbf{Y}'-[(\mathbf{Y}')^{-1}\ud\mathbf{Y}']^2}.\nonumber
\end{equation}
The third differential is
\begin{equation}\label{third_diff}
\ud^3 L(\mathbf{M})=-\frac{2}{ N\beta}{\rm
Tr}[(\mathbf{G}\ud\mathbf{G}^{-1})^3]+\frac{1}{ N\beta } {\rm
Tr}[\ud^3\ln(\mathbf{Y}')]-\mathbf{W}^{-1} (3\ud^2 \mathbf{Y} \ud \mathbf{Y}+\mathbf{Y}
 \ud^3 \mathbf{Y}),
\end{equation}
with
\begin{equation}
{\rm Tr}[\ud^3\ln(\mathbf{Y}')] = {\rm Tr}\{(\mathbf{Y}')^{-1}\ud^3\mathbf{Y}'-3(\mathbf{Y}')^{-1}\ud\mathbf{Y}'(\mathbf{Y}')^{-1}\ud^2\mathbf{Y}'
+2[(\mathbf{Y}')^{-1}\ud\mathbf{Y}']^3\}.\nonumber
\end{equation}
The forth differential, correspondingly, reads
\begin{equation}\label{forth_diff}
\ud^4 L(\mathbf{M})=\frac{6}{ N\beta}{\rm
Tr}[(\mathbf{G}\ud\mathbf{G}^{-1})^4]+\frac{1}{ N\beta } {\rm
Tr}[\ud^4\ln(\mathbf{Y}')]-\mathbf{W}^{-1} (4\ud^3 \mathbf{Y} \ud \mathbf{Y}+3(\ud^2 \mathbf{Y})^2+\mathbf{Y}
 \ud^4 \mathbf{Y}),
\end{equation}
with
\begin{align}
\begin{split}
{\rm Tr}[\ud^4\ln(\mathbf{Y}')] = {\rm Tr}\{(\mathbf{Y}')^{-1}\ud^4\mathbf{Y}'&-4(\mathbf{Y}')^{-1}\ud\mathbf{Y}'(\mathbf{Y}')^{-1}\ud^3\mathbf{Y}'-3[(\mathbf{Y}')^{-1}\ud^2\mathbf{Y}']^2\\
&+12[(\mathbf{Y}')^{-1}\ud\mathbf{Y}']^2(\mathbf{Y}')^{-1}\ud^2\mathbf{Y}'
-6[(\mathbf{Y}')^{-1}\ud\mathbf{Y}']^4\}.
\end{split}\nonumber
\end{align}

As we can see, all the equations beyond the second differential involve the interaction term $\mathbf{U}$ only in the renormalized form, through $\mathbf{W}$. Now, we require the vanishing of the differentials at the saddle point and recall that $\mathbf{Y}(\mathbf{M}_{\rm sp})=0$. All the derivatives below are also evaluated at the point $\mathbf{M}=\mathbf{M}_{\rm sp}$, and we use Eq.\;(\ref{Gamma_def}). 
The first differential yields
\begin{equation}\label{M_sp_exact}
\mathbf{U}^{-1}\mathbf{M}_{\rm sp}\ud\mathbf{M} =\Gamma^{(1)}\ud\mathbf{M}-\frac{1}{ N\beta } {\rm Tr}[\mathbf{Y}''\ud\mathbf{M}],
\end{equation}
where the last term describes the contribution to the mean field, beyond the Hartree-Fock approximation. The second differential yields
\begin{equation}\label{W_exact}
\mathbf{W}^{-1} \ud\mathbf{M}^2 = \mathbf{U}^{-1}\ud \mathbf{M}^2-\Gamma^{(2)}\ud\mathbf{M}^2+\frac{1}{ N\beta }{\rm Tr}[\mathbf{Y}'''\ud\mathbf{M}^2-(\mathbf{Y}''\ud\mathbf{M})^2],
\end{equation}
where the term with the trace describes the contribution to the effective interaction beyond RPA.
Eq.\;(\ref{M_sp_exact}) constrains the value of $\mathbf{M}_{\rm sp}$, which is determined for a given value of $\mathbf{Y}''(\mathbf{M}_{\rm sp})$, while Eq.\;(\ref{W_exact}) relates the derivatives of the variable transformation to  $\mathbf{W}^{-1}$. While the relation is exact, in the simplest approximation the term proportional to $\mathbf{Y}'''$ gives rise to a Maki-Thompson\cite{Maki,Thompson} type correction to the effective interaction and the term proportional to the square of $\mathbf{Y}''$ gives rise to an Aslamazov-Larkin correction.\cite{Aslamazov_Larkin} In order to have a controllable connection between the saddle-point approximation and the exact case, we introduce a parameter $l$ in front of ${\rm Tr}[\ln(-\mathbf{G}^{-1})]$ for all differentials beyond the second order, which corresponds to the action
\begin{equation}\label{HS_action_l}
	{\cal S}_l(\mathbf{M})=\frac{N\hbar\beta}{ 2 }[\mathbf{U}^{-1}\mathbf{M}
	\mathbf{M}-(1-l)\Gamma^{(2)}(\mathbf{M}-\mathbf{M}_{\rm sp})^2-2(1-l)\Gamma^{(1)}(\mathbf{M}-\mathbf{M}_{\rm sp})]-\hbar{\rm Tr}[\ln(-\mathbf{G}^{-l}(\mathbf{M})\mathbf{G}^{l-1}(\mathbf{M}_{\rm sp}))].
\end{equation}
Setting the parameter $l=0$ corresponds to the saddle-point approximation, while $l=1$ corresponds to the exact case. Evaluating Eq.\;(\ref{third_diff}) and Eq.\;(\ref{forth_diff}) at  $\mathbf{M}_{\rm sp}$ and using that  $\mathbf{Y}'(\mathbf{M}_{\rm sp}) = \mathbf{I}$, we obtain
\begin{equation}
3 \mathbf{W}^{-1}\ud \mathbf{M}\mathbf{Y}''\ud \mathbf{M}^2 = -l\Gamma^{(3)}\ud \mathbf{M}^3+\frac{1}{ N\beta }{\rm Tr}[\mathbf{Y}^{(4)}\ud\mathbf{M}^3-3\mathbf{Y}''\ud\mathbf{M}\mathbf{Y}'''\ud\mathbf{M}^2+2(\mathbf{Y}''\ud\mathbf{M})^3]
\end{equation}
and
\begin{equation}
\mathbf{W}^{-1} (4\ud \mathbf{M}\mathbf{Y}'''\ud \mathbf{M}^3+3(\mathbf{Y}''\ud \mathbf{M}^2)^2)= -l\Gamma^{(4)}\ud \mathbf{M}^4+\frac{1}{ N\beta } {\rm Tr}[\ud^4\ln(\mathbf{Y}')],
\end{equation}
where
\begin{equation}\label{d4_log}
{\rm Tr}[\ud^4\ln(\mathbf{Y}')] = {\rm Tr}[\mathbf{Y}^{(5)}\ud\mathbf{M}^4-4\mathbf{Y}''\ud\mathbf{M}\mathbf{Y}^{(4)}\ud\mathbf{M}^3-3(\mathbf{Y}'''\ud\mathbf{M}^2)^2+12(\mathbf{Y}''\ud\mathbf{M})^2\mathbf{Y}'''\ud\mathbf{M}^2
-6(\mathbf{Y}''\ud\mathbf{M})^4].
\end{equation}

In general, we can parametrize the variable $\mathbf{M}$ as $\mathbf{M}(\mathbf{X})$, such that $\mathbf{M}(0)=\mathbf{M}_{\rm sp}$ and $\mathbf{M}'(0)=\mathbf{I}$ and consider $\mathbf{X}$ to be an independent variable, defining $\tilde{L}(\mathbf{X})\equiv L(\mathbf{M}(\mathbf{X}))$. Clearly, the equation $\ud \tilde{L}(0)=0$ that constrains the value of $\mathbf{M}(0)$ does not depend on such parametrization. Moreover, all the $k$ equations $\ud \tilde{L}(0)=0$, $\ud^2 \tilde{L}(0)=0$, ..., $\ud^k \tilde{L}(0)=0$ are invariant with respect to the parametrization, i.e. they do not depend on the parametrization used for their derivation. As we have already seen, the equation $\ud \tilde{L}(0)=0$ constrains the value of $\mathbf{M}(0)$. Analogously, the equation $\ud^k \tilde{L}(0)=0$ constrains the value of the derivative $\mathbf{Y}^{(k-1)}(\mathbf{M}_{\rm sp})$. However, the $k$ equations do not form a closed system for the derivatives of $\mathbf{Y}$ at the saddle point, because the equations $\ud^{k-1} \tilde{L}(0)=0$ and $\ud^{k} \tilde{L}(0)=0$ involve derivatives $\mathbf{Y}^{(k)}(\mathbf{M}_{\rm sp})$ and $\mathbf{Y}^{(k+1)}(\mathbf{M}_{\rm sp})$ for which there are no direct constraints within the system. These higher-order terms appear due to the presence of $\ud^{k-1} {\rm Tr}[\ln(\mathbf{Y}')]$ and $\ud^{k} {\rm Tr}[\ln(\mathbf{Y}')]$ in the equations. Once their values are specified, the system of equations becomes closed. The asymptotic value of a derivative directly follows from the equation of the corresponding order, when $l \rightarrow 0$ (or $\left\|\mathbf{W}\right\|\rightarrow 0$):
\begin{equation}
(k+1)\mathbf{W}^{-1} \ud \mathbf{M}\mathbf{Y}^{(k)}\ud\mathbf{M}^{k} \rightarrow -l\Gamma^{(k+1)}\ud\mathbf{M}^{k+1}+\frac{1}{ N\beta }{\rm Tr}[\mathbf{Y}^{(k+2)}\ud\mathbf{M}^{k+1}].
\end{equation}
However, for a generic system, the derivatives are not known. Therefore, for a non-perturbative approximation, the simplest choice is to consider coordinate transformations that satisfy the system of $k$ equations and have $\mathbf{Y}^{(k)}(\mathbf{M}_{\rm sp})=0$ and $\mathbf{Y}^{(k+1)}(\mathbf{M}_{\rm sp})=0$ at the saddle point (the constraints are specific to the particular parametrization, i.e. choosing $\mathbf{M}$ as an independent variable). These constraints, in general, are not consistent with the higher-order equations, implying that $\ud^{m} L(\mathbf{M}_{\rm sp})\neq0$ for $m>k$. This means that after such coordinate transformation, the saddle point is not strictly quadratic,
\begin{equation}
{\cal S}_{\rm sp}(\mathbf{Y})={\cal S}_{\rm
sp}+\frac{N\hbar\beta}{2}\mathbf{W}^{-1}\mathbf{Y}\mathbf{Y} +{\cal
O}(\mathbf{Y}^{k+1}).
\end{equation}

With the above choice of the coordinate transformations, the $k$ equations form a closed system of nonlinear tensorial equations that completely determine the partition function $Z[\mathbf{h}]$. Such approximation is a generalization of the saddle-point approximation, i.e. $k=2$ corresponds to the saddle-point approximation, while $k=3$ introduces corrections to the saddle point and the renormalized interaction term $\mathbf{W}^{-1}$. However, as we will see in Section \ref{one_dimensional_integrals}, in the case of scalar functions the constraints are not very suitable at the phase-transition point. Therefore, we are going to consider a different approach, namely, the leading-order approximation for the derivatives (linear in $l$), assuming that the asymptotic behavior holds, 
\begin{equation}\label{asymptotic_proxy}
(k+1)\mathbf{W}^{-1} \ud \mathbf{M}\mathbf{Y}^{(k)}\ud\mathbf{M}^{k} = -\Gamma^{(k+1)}\ud\mathbf{M}^{k+1}+\frac{1}{ N\beta }{\rm Tr}[\mathbf{Y}^{(k+2)}\ud\mathbf{M}^{k+1}].
\end{equation}
The approach is independent from a particular choice of parametrization $\mathbf{M}(\mathbf{X})$. Isolating $\mathbf{Y}^{(k)}$ leads to
\begin{align}
\begin{split}
Y^{(k)\; (\mathbf{p},r)}_{(\mathbf{p}^{1},r^{1})\ldots (\mathbf{p}^{k},r^{k})}\ud M^{(\mathbf{p}^{k},r^{k})}\ldots &\ud M^{(\mathbf{p}^{1},r^{1})} = -\frac{1}{(k+1)}\Gamma^{(k+1)}_{(\mathbf{q},s),(\mathbf{p}^{1},r^{1})\ldots (\mathbf{p}^{k},r^{k})}\ud M^{(\mathbf{p}^{k},r^{k})}\ldots \ud M^{(\mathbf{p}^{1},r^{1})}W^{(\mathbf{q},s),(\mathbf{p},r)}\\
+&\frac{1}{(k+1)(N\beta)}Y^{(k+2)\; (\mathbf{q}^{1},s^{1})}_{(\mathbf{q}^{1},s^{1}),(\mathbf{q},s),(\mathbf{p}^{1},r^{1})\ldots (\mathbf{p}^{k},r^{k})}\ud M^{(\mathbf{p}^{k},r^{k})}\ldots \ud M^{(\mathbf{p}^{1},r^{1})}W^{(\mathbf{q},s),(\mathbf{p},r)},
\end{split}
\end{align}
which can be verified by a direct substitution of $\mathbf{Y}^{(k)}$ into Eq.\;(\ref{asymptotic_proxy}). Taking its trace yields
\begin{equation}
\frac{1}{N\beta }{\rm Tr}[\mathbf{Y}^{(k)}\ud\mathbf{M}^{k-1}] = -\frac{1}{(k+1) N\beta }{\rm Tr}[\Gamma^{(k+1)}\ud\mathbf{M}^{k-1}\mathbf{W}]+\frac{1}{(k+1) (N\beta)^2 }{\rm Tr}^2[\mathbf{Y}^{(k+2)}\ud\mathbf{M}^{k}\mathbf{W}].
\end{equation}
The recursive relation leads to either even $k$-order in $\mathbf{Y}^{(k)}$,
\begin{equation}\label{second_diff_leading}
\frac{1}{ N\beta } {\rm Tr}[\mathbf{Y}''\ud\mathbf{M}]=-\sum_{k=1} \frac{1}{(2k+1)!! (N\beta)^{k} }{\rm Tr}^{k}[\Gamma^{(2k+1)}\ud\mathbf{M}\mathbf{W}^{k}],
\end{equation}
or to odd
\begin{equation}\label{third_diff_leading}
\frac{1}{ N\beta }{\rm Tr}[\mathbf{Y}'''\ud\mathbf{M}^2]=-\sum_{k=1} \frac{2}{(2k+2)!! (N\beta)^{k} }{\rm Tr}^{k}[\Gamma^{(2k+2)}\ud\mathbf{M}^2\mathbf{W}^{k}],
\end{equation}
where ${\rm Tr}^{k}$ denotes contraction over $k$ pairs of indices. For example, for $k=1$ the ${\rm Tr}$ contraction is performed over $(\mathbf{p}_2,r_2)$ indices
\begin{equation}
\frac{1}{3 N\beta }{\rm Tr}[\Gamma^{(3)}\ud\mathbf{M}\mathbf{W}]=\frac{1}{3 N\beta }\Gamma^{(3)}_{(\mathbf{p}_2,r_2), (\mathbf{p}_1,r_1), (\mathbf{p},r)}\ud M^{(\mathbf{p},r)} 
W^{(\mathbf{p}_1,r_1),(\mathbf{p}_2,r_2)},
\end{equation}
while for $k=2$ the ${\rm Tr}^2$ contraction is performed over $(\mathbf{p}_4,r_4),(\mathbf{p}_3,r_3)$ indices,
\begin{equation}
 \frac{1}{5!! (N\beta)^{2} }{\rm Tr}^{2}[\Gamma^{(5)}\ud\mathbf{M}\mathbf{W}^{2}] = \frac{1}{5!! (N\beta)^{2} }\Gamma^{(5)}_{(\mathbf{p}_4,r_4),(\mathbf{p}_3,r_3),(\mathbf{p}_2,r_2), (\mathbf{p}_1,r_1), (\mathbf{p},r)}\ud M^{(\mathbf{p},r)}W^{(\mathbf{p}_1,r_1),(\mathbf{p}_3,r_3)}W^{(\mathbf{p}_2,r_2),(\mathbf{p}_4,r_4)}.
\end{equation}
Therefore, in the leading-order approximation, after substituting Eq.\;(\ref{second_diff_leading}) into  Eq.\;(\ref{M_sp_exact}) one finds
\begin{equation}\label{M_sp_leading}
\mathbf{U}^{-1}\mathbf{M}_{\rm sp}\ud\mathbf{M} =\Gamma^{(1)}\ud\mathbf{M}+\sum_{k=1} \frac{1}{(2k+1)!! (N\beta)^{k} }{\rm Tr}^{k}[\Gamma^{(2k+1)}\ud\mathbf{M}\mathbf{W}^{k}],
\end{equation}
and equivalently, upon substituting Eq.\;(\ref{third_diff_leading}) into Eq.\;(\ref{W_exact}) and neglecting higher-order non-linear terms, 
\begin{equation}\label{W_leading}
\mathbf{W}^{-1} \ud\mathbf{M}^2 = \mathbf{U}^{-1}\ud \mathbf{M}^2-\Gamma^{(2)}\ud\mathbf{M}^2-\sum_{k=1} \frac{2}{(2k+2)!! (N\beta)^{k} }{\rm Tr}^{k}[\Gamma^{(2k+2)}\ud\mathbf{M}^2\mathbf{W}^{k}].
\end{equation}
Eq.\;(\ref{M_sp_leading}) and Eq.\;(\ref{W_leading}) represent, respectively, the Hartree-Fock approximation and the RPA results, with the first-order corrections in the fermionic loops $\Gamma^{(k)}$ with $k>2$ (more than two propagators). In particular, the approximation includes Maki-Thompson\cite{Maki,Thompson} correction to the effective interaction. However, it implies that the contribution from two (and higher) fermionic loop diagrams, such as FLEX\cite{Bickers_Scalapino_White,Bickers_Scalapino} or Aslamazov-Larkin corrections,\cite{Aslamazov_Larkin} is not included in the approximation. 

\section{Correlation functions and linear response}
Now that the formalism has been developed, we apply it to the calculation of correlation functions, which can be promptly evaluated by taking functional derivatives with respect to the source $\mathbf{h}$. In general, the following exact relation holds,
\begin{equation}
Z=\exp\left(-\frac{N\beta}{ 2
}\mathbf{U}^{-1}\mathbf{M}_{\rm sp}\mathbf{M}_{\rm sp}+{\rm
Tr}[\ln(-\mathbf{G}^{-1}(\mathbf{M}_{\rm sp}))]-\frac{1}{2}{\rm Tr}[\ln(\mathbf{U}\mathbf{W}^{-1})]\right),
\end{equation}
which in combination with a closed form solution for $\mathbf{M}_{\rm sp}$ and $\mathbf{W}^{-1}$, determines $Z[\mathbf{h}]$ (recall that $\mathbf{M}_{\rm sp}$ and $W$ depend on $\Gamma$, which on its turn depends on $\mathbf{h}$, see Eq.\;(\ref{Greens_function_with_h})). Instead of $Z$ itself, it is more convenient to use $\ln(Z)$, which is
\begin{equation}\label{ln_z}
\ln(Z)=-\frac{N\beta}{ 2
}\mathbf{U}^{-1}\mathbf{M}_{\rm sp}\mathbf{M}_{\rm sp}+{\rm
Tr}[\ln(-\mathbf{G}^{-1}(\mathbf{M}_{\rm sp}))]-\frac{1}{2}{\rm Tr}[\ln(\mathbf{U}\mathbf{W}^{-1})].
\end{equation}

Now we return to Eq.\;(\ref{delta_gamma}), where $\delta_{\mathbf{h}}  \Gamma^{(n)}$ denotes variation with respect to $\mathbf{h}$ only, with $\mathbf{M}_{\rm sp}$ fixed, which is calculated by using the identity $\delta_{\mathbf{h}} \mathbf{G}^{-1}=\delta\mathbf{h} $. In particular,
\begin{equation}\label{delta_ln_g}
\delta_{\mathbf{h}}{\rm Tr}[\ln(-\mathbf{G}^{-1}(\mathbf{M}_{\rm sp}))]={\rm Tr}[\mathbf{G}(\mathbf{M}_{\rm sp})\delta\mathbf{h}].
\end{equation}
Differentiating Eq.\;(\ref{ln_z}) and using Eq.\;(\ref{delta_ln_g}),we find that the first variation of $\ln(Z)$ with respect to $\delta\mathbf{h}$ reads
\begin{equation}
\delta  \ln(Z[\mathbf{h}])=-N\beta\mathbf{U}^{-1}\mathbf{M}_{\rm sp}\delta \mathbf{M}_{\rm sp}+N\beta\Gamma^{(1)}\delta \mathbf{M}_{\rm sp}+{\rm Tr}[\mathbf{G}(\mathbf{M}_{\rm sp})\delta\mathbf{h}]-\frac{1}{2}{\rm Tr}[\mathbf{W} \delta\mathbf{W}^{-1}]
\end{equation}
and can be simplified by using Eq.\;(\ref{M_sp_exact}) with $\ud\mathbf{M}$ substituted by $\delta \mathbf{M}_{\rm sp}$, 
\begin{equation}
\mathbf{U}^{-1}\mathbf{M}_{\rm sp}\delta \mathbf{M}_{\rm sp} =\Gamma^{(1)}\delta \mathbf{M}_{\rm sp}-\frac{1}{ N\beta }{\rm Tr}[\mathbf{Y}''(\mathbf{M}_{\rm sp})\delta \mathbf{M}_{\rm sp}]
\end{equation}
to yield
\begin{equation}
\delta \ln(Z[\mathbf{h}])={\rm
Tr}[\mathbf{G}(\mathbf{M}_{\rm sp})\delta\mathbf{h}]+{\rm Tr}[\mathbf{Y}''(\mathbf{M}_{\rm sp})\delta \mathbf{M}_{\rm sp}]-\frac{1}{2}{\rm Tr}[\mathbf{W} \delta\mathbf{W}^{-1}].
\end{equation}
Moreover, in case $\mathbf{h}$ is related to a vector $\mathbf{B}$ as
\begin{equation}
\mathbf{h}_{\mathbf{q},\mathbf{q}'}=\sum_{\mathbf{p},r}\mathbf{R}_{\mathbf{q},\mathbf{q}';\mathbf{p},r}B^{\mathbf{p},r},
\end{equation}
the variation of $\Gamma^{(n)}$ simplifies to
\begin{equation}\label{delta_gamma_b}
\delta  \Gamma^{(n)}=  \Gamma^{(n+1)}(\delta \mathbf{B}+ \delta\mathbf{M}_{\rm sp}).
\end{equation} 

Therefore, in the leading-order approximation, the variation of Eq.\;(\ref{M_sp_leading}) reads
\begin{equation}
\mathbf{U}^{-1}\delta\mathbf{M}_{\rm sp}\ud\mathbf{M} =\Gamma^{(2)}(\delta \mathbf{B}+ \delta\mathbf{M}_{\rm sp})\ud\mathbf{M}+\sum_{k=1} \frac{1}{(2k+1)!! (N\beta)^{k} }{\rm Tr}^{k}[\Gamma^{(2k+2)}(\delta \mathbf{B}+ \delta\mathbf{M}_{\rm sp})\ud\mathbf{M}\mathbf{W}^{k}] + \cal{O}(\delta \mathbf{W}),
\end{equation}
where we used Eq.\;(\ref{delta_gamma_b}) and where $\delta \mathbf{W}$ is obtained by the variation of Eq.\;(\ref{W_leading}),
\begin{equation}
\delta\mathbf{W}^{-1} \ud\mathbf{M}^2 = -\Gamma^{(3)}(\delta \mathbf{B}+ \delta\mathbf{M}_{\rm sp})\ud\mathbf{M}^2-\sum_{k=1} \frac{2}{(2k+2)!! (N\beta)^{k} }{\rm Tr}^{k}[\Gamma^{(2k+3)}(\delta \mathbf{B}+ \delta\mathbf{M}_{\rm sp})\ud\mathbf{M}^2\mathbf{W}^{k}] + \cal{O}(\delta \mathbf{W}).
\end{equation}

Introducing, for convenience, a symmetric tensor
\begin{equation}\label{v_def}
\mathbf{V}\ud\mathbf{M}^2 = \Gamma^{(2)}\ud\mathbf{M}^2+\sum_{k=1} \frac{1}{(2k+1)!! (N\beta)^{k} }{\rm Tr}^{k}[\Gamma^{(2k+2)}\ud\mathbf{M}^2\mathbf{W}^{k}]
\end{equation}
and neglecting $\cal{O}(\delta \mathbf{W})$ yields
\begin{equation}
\mathbf{U}^{-1}\delta\mathbf{M}_{\rm sp} =\mathbf{V}(\delta \mathbf{B}+ \delta\mathbf{M}_{\rm sp}),
\end{equation}
which leads to 
\begin{equation}
\delta\mathbf{M}_{\rm sp} =(\mathbf{I}-\mathbf{U}\mathbf{V})^{-1}\mathbf{U}\mathbf{V}\delta \mathbf{B},
\end{equation}
or, equivalently,
\begin{equation}
\delta\mathbf{M}_{\rm sp} =[\mathbf{I}-\mathbf{W}(\mathbf{W}^{-1}-\mathbf{U}^{-1}+\mathbf{V})]^{-1}\mathbf{W}\mathbf{V}\delta \mathbf{B}.
\end{equation}
Clearly, the instability appears when $\mathbf{W}(\mathbf{W}^{-1}-\mathbf{U}^{-1}+\mathbf{V})$ acquires an eigenvalue equal or larger than one, i.e.
\begin{equation}\label{instability_condition}
\mathbf{W}^{-1}\delta\mathbf{M}_{\rm sp}\ud\mathbf{M} = \sum_{k=1} \left(\frac{1}{(2k+1)!!}-\frac{2}{(2k+2)!!}\right) \frac{1}{(N\beta)^{k} }{\rm Tr}^{k}[\Gamma^{(2k+2)}\delta\mathbf{M}_{\rm sp}\ud\mathbf{M}\mathbf{W}^{k}],
\end{equation} 
where we substituted Eq.\;(\ref{v_def}) and Eq.\;(\ref{W_leading}) into the eigenvalue equation. The presence of the non-trivial prefactors in Eq.\;(\ref{instability_condition}) suggests that in the vicinity of a phase transition the series resummation might yield a different scaling dependence than that for an ordinary geometric series or a mean-field approach. In Ref. [\onlinecite{EPL}], a fermionic system is investigated in the presence of anisotropic Zeeman terms within the saddle-point approximation. In that case, the right-hand side of Eq.\;(\ref{instability_condition}) vanishes and $\mathbf{W}$ is the effective RPA interaction, corresponding to the generalized susceptibility, the divergence of which signals the transition into a spin-charge-density wave phase. With the formalism developed here, the corrections to this result could be promptly evaluated. It is important to emphasize that contrary to a spread assumption,\cite{Dahm_Scalapino,Chubukov_Pines_Schmalian,Esirgen_Schuttler_Grober_Evertz} the divergence of the effective interactions at the point of the phase transition is an artifact of the RPA approximation. In general, this does not hold when non trivial higher order terms are present. This conclusion directly follows from the definition of $\mathbf{W}$, implying that the partition function $Z$ is proportional to $\det[\mathbf{W}]^{1/2}$, assuming finite $\mathbf{M}_{\rm sp}$. Therefore, as long as the partition function remains finite at the point of the phase transition, so does the effective interaction $\mathbf{W}$.  

\section{Cooper pair instability}
\label{Cooper_instability}
The previous section describes a transition into a spin-charge-density wave phase as an instability of the saddle-point value $\mathbf{M}_{\rm sp}$. In the present section, we are considering an instability towards Cooper pair formation, which is signalled by the divergence of the corresponding correlation function, defined in Eq.\;(\ref{cooper_corrfunc}). 
Namely, the correlation function is generated when adding a source term of the BCS form Eq.\;(\ref{cooper_source}) and differentiating with respect to  $\Delta$ and $\Delta^{*}$. This term can be interpreted as a pairing interaction term
\begin{equation}
\sum_{\mathbf{q}_1,\mathbf{q}_2}\rho^{a}_{\mathbf{q}_1,\mathbf{q}_2} a^{\dag}_{\mathbf{q}_1,\uparrow}a^{\dag}_{-\mathbf{q}_1,\downarrow}a_{\mathbf{q}_2,\uparrow}a_{-\mathbf{q}_2,\downarrow}
\end{equation}
where $\rho^{a}_{\mathbf{q}_1,\mathbf{q}_2} =\Delta^{*}_{\mathbf{q}_1,-\mathbf{q}_1}\Delta_{\mathbf{q}_2,-\mathbf{q}_2}$. 
This is a specific case of the more general interaction defined earlier by Eq.\;(\ref{u_total_inbound}), which depends on the total inbound momentum, 
\begin{equation}
\mathbf{U}_{\mathbf{q}_1+\mathbf{q}_1'}^{(\mathbf{p},r),(\mathbf{p}',r')}=\delta_{\mathbf{q}_1+\mathbf{q}_1',0}\delta_{\mathbf{q}_1-\mathbf{q}_2,\mathbf{p}}\delta_{\mathbf{p}+\mathbf{p}',0}\Delta^{*}_{\mathbf{q}_1,\mathbf{q}_1' }\Delta_{\mathbf{q}_2,\mathbf{q}_2'}\mathbf{Q}^{r,r'},
\end{equation}
with some matrix $\mathbf{Q}^{r,r'}$. In particular, the term is equivalently represented as a result of integrating out the auxiliary fields $h_{+}$ and $h_{-}$, defined for the $A$ sublattice as
\begin{equation}
\sum_{\mathbf{q}_1,\mathbf{q}_2}(\psi^{\dag}_{\mathbf{q}_1}\mathbf{P}^{a}_{+}\psi_{-\mathbf{q}_2}h^{a}_{\mathbf{q}_1,-\mathbf{q}_2,+}+\psi^{\dag}_{-\mathbf{q}_1}\mathbf{P}^{a}_{-}\psi_{\mathbf{q}_2}h^{a}_{-\mathbf{q}_1,\mathbf{q}_2,-}-(\rho^{a }_{\mathbf{q}_1,\mathbf{q}_2})^{-1}h^{a}_{\mathbf{q}_1,-\mathbf{q}_2,+}h^{a}_{-\mathbf{q}_1,\mathbf{q}_2,-}),
\end{equation}
where $\mathbf{P}^{a}_{+}$ and $\mathbf{P}^{a}_{-}$ are such that $\psi^{\dag}_{\mathbf{q}_1}\mathbf{P}^{a}_{+}\psi_{-\mathbf{q}_2}= a^{\dag}_{\mathbf{q}_1,\uparrow}a_{-\mathbf{q}_2,\downarrow}$ and $\psi^{\dag}_{-\mathbf{q}_1}\mathbf{P}^{a}_{-}\psi_{\mathbf{q}_2}= a^{\dag}_{-\mathbf{q}_1,\downarrow}a_{\mathbf{q}_2,\uparrow}$ and analogously for the $B$ sublattice.
This allows us to investigate the stability of the system with respect to the presence of the pairing interaction. In
general, the divergence of the derivatives of $\ln(Z[\mathbf{h}])$ signal a phase transition. Our partition function is determined by $\mathbf{M}_{\rm sp}$ and $\mathbf{W}$. For spin- and charge-density wave formation, the instability is signaled by a divergence of the derivative of $\mathbf{M}_{\rm sp}$, whereas pairing arises from a divergence in the derivative of $\mathbf{W}$. Below, we demonstrate that the divergence of the second derivative of $\ln(Z[\mathbf{h}])$ is caused by an instability of the renormalized interaction $\mathbf{W}$ towards acquiring a dependence on the inbound momentum, which corresponds to the emergence of the pairing interaction of the form
\begin{equation}
(\mathbf{W}^{-1})^{\mathbf{q}+\mathbf{q}'}_{(\mathbf{p},r),(\mathbf{p}',r')}M^{(\mathbf{p},r)}_{\mathbf{q}}
M^{(\mathbf{p}',r')}_{\mathbf{q}'}.
\end{equation}
This instability implies that there is no transformation $\mathbf{Y}(\mathbf{M})$ satisfying Eq.\;(\ref{SP_action}) for a system in a superconducting state. 

Let us investigate the response of the system to an infinitesimally small interaction that depends on the total inbound momentum. Because $\ln(Z[\mathbf{h}])$ has the term ${\rm Tr}[\ln(\mathbf{W}^{-1})]$, the second variation of $\ln(Z[\mathbf{h}])$ has the term
\begin{equation}
\delta_\rho{\rm Tr}[\ln(\mathbf{W}^{-1})] =  \delta_{\mathbf{h}}\delta_{\mathbf{h}'}{\rm Tr}[\ln(\mathbf{W}^{-1})].
\end{equation}
The first variation yields
\begin{equation}
\delta{\rm Tr}[\ln(\mathbf{W}^{-1})] =  {\rm Tr}[\mathbf{W} \delta\mathbf{W}^{-1}],
\end{equation}
which after substitution of Eq.\;(\ref{W_leading}) becomes 
\begin{equation}
{\rm Tr}[\mathbf{W} \delta\mathbf{W}^{-1}] = -{\rm Tr}[\mathbf{W} \delta\Gamma^{(2)}]-\sum_{k=1} \frac{2}{(2k+2)!! (N\beta)^{k} }{\rm Tr}^{k+1}[\mathbf{W}\delta(\Gamma^{(2k+2)}\mathbf{W}^{k})].
\end{equation}
The second variation yields
\begin{equation}
\delta {\rm Tr}[\mathbf{W} \delta\mathbf{W}^{-1}] = {\rm Tr}[\mathbf{W} \delta^2\mathbf{W}^{-1}] - {\rm Tr}[(\mathbf{W} \delta\mathbf{W}^{-1})^2].
\end{equation}

In particular, the expression for ${\rm Tr}[\mathbf{W} \delta^2\mathbf{W}^{-1}]$ is given by 
\begin{equation} \label{Stefan}
{\rm Tr}[\mathbf{W} \delta^2\mathbf{W}^{-1}] = -{\rm Tr}[\mathbf{W} \delta_{\mathbf{h}}\delta_{\mathbf{h}'}\Gamma^{(2)}]-\sum_{k=2} \frac{2}{(2k)!! (N\beta)^{k-1} }{\rm Tr}^{k}[\delta_{\mathbf{h}}\delta_{\mathbf{h}'}\Gamma^{(2k)}\mathbf{W}^{k}] + \cal{O}(\delta \mathbf{W}) + \cal{O}(\delta \mathbf{M}_{\rm sp}).
\end{equation}
The first term has $\delta_{\mathbf{h}}\delta_{\mathbf{h}'}\Gamma^{(2)}$, more explicitly
\begin{equation}
{\rm Tr}[\mathbf{W} \delta_{\mathbf{h}}\delta_{\mathbf{h}'}\Gamma^{(2)}] = W^{(\mathbf{p},r),(\mathbf{p}',r')}\delta_{\mathbf{h}}\delta_{\mathbf{h}'}\Gamma^{(2)}_{(\mathbf{p},r),(\mathbf{p}',r')},
\end{equation}
which is nothing but 
\begin{equation}
{\rm Tr}[\mathbf{W} \delta^2_{\mathbf{h}}\Gamma^{(2)}] = \frac{-2}{ \beta N }W^{(\mathbf{p},r),(\mathbf{p}',r')}\sum_{\mathbf{q},\mathbf{q}',\mathbf{h},\mathbf{h}'}{\rm
Tr}[\delta_{\mathbf{h}}\mathbf{RG}_{\mathbf{q},\mathbf{q'};\mathbf{p},r}\delta_{\mathbf{h}'}\mathbf{RG}_{\mathbf{q}',\mathbf{q};\mathbf{p}',r'}+
\mathbf{RG}_{\mathbf{q},\mathbf{q'};\mathbf{p},r}\delta_{\mathbf{h}}\delta_{\mathbf{h}'}\mathbf{RG}_{\mathbf{q}',\mathbf{q};\mathbf{p}',r'}],
\end{equation}
where we used Eq.\;(\ref{RG_def}). Recalling that $\delta_{\mathbf{h}}\mathbf{G}= -\mathbf{G} \delta \mathbf{h}\mathbf{G}$, the first term is
\begin{equation}
\frac{-2}{ \beta N }W^{(\mathbf{p},r),(\mathbf{p}',r')}\sum_{\mathbf{q},...,\mathbf{q}_{2},\mathbf{q}',...,\mathbf{q}_{2}',\mathbf{h},\mathbf{h}'}{\rm
Tr}[\mathbf{RG}_{\mathbf{q},\mathbf{q}_1;\mathbf{p},r}\delta\mathbf{h}_{\mathbf{q}_1,\mathbf{q}_{1}'}\mathbf{G}_{\mathbf{q}_{1}',\mathbf{q}'}\mathbf{RG}_{\mathbf{q}',\mathbf{q}_2';\mathbf{p}',r'}\delta\mathbf{h}'_{\mathbf{q}_2',\mathbf{q}_{2}}\mathbf{G}_{\mathbf{q}_{2},\mathbf{q}}].
\end{equation}
The higher-order  terms corresponding to the 'wheel' diagram have an analogous form,
\begin{align}
\begin{split}
\sum_{\mathbf{q},...,\mathbf{q}_{k+1},\mathbf{q}',...,\mathbf{q}_{k+1}',\mathbf{h},\mathbf{h}'} &\frac{-(2k)!!}{ (N\beta)^{k} }{\rm Tr}[\delta\mathbf{h}_{\mathbf{q}',\mathbf{q}_{k+1}}\mathbf{G}_{\mathbf{q}_{k+1},\mathbf{q}_k}\mathbf{RG}_{\mathbf{q}_{k},\mathbf{q}_{k-1};\mathbf{p}_k,r_k}...\mathbf{RG}_{\mathbf{q}_1,\mathbf{q};\mathbf{p}_1,r_1}\\
&\times \delta\mathbf{h}'_{\mathbf{q},\mathbf{q}_{k+1}'}\mathbf{G}_{\mathbf{q}_{k+1}',\mathbf{q}_{k}'}\mathbf{RG}_{\mathbf{q}_{k}',\mathbf{q}_{k-1}';\mathbf{p}_k',r_k'}...\mathbf{RG}_{\mathbf{q}_{1}',\mathbf{q}';\mathbf{p}_1',r_1'}]
 W^{(\mathbf{p}_1,r_1),(\mathbf{p}_1',r_1')}...W^{(\mathbf{p}_k,r_k),(\mathbf{p}_k',r_k')},
\end{split}
\end{align}
where the prefactor $(2k)!!$ reflects the symmetry of the corresponding diagram. 
The variations are such that for any matrix $\mathbf{A}$ and $\mathbf{B}$ of appropriate dimensions
\begin{equation}
\frac{1}{ N\beta }\sum_{\mathbf{q}_1,\mathbf{q}_1'}{\rm Tr}[\mathbf{A}\delta\mathbf{h}(\mathbf{q}_1,\mathbf{q}_1')\mathbf{B}\delta\mathbf{h}'(\mathbf{q}_1',\mathbf{q}_1)] = \frac{1}{ N\beta }{\rm Tr}[\mathbf{A}\Delta\mathbf{B}^T\Delta^{\dag}].
\end{equation}

\begin{figure}
\begin{tikzpicture}
	\begin{feynman}
			\vertex (p) {\(q\)};
		\vertex [right=of p] (p1);
		\vertex [right=of p1] (p2); 
		\vertex [right=of p2] (p3);
		\vertex [right=of p3] (p4);
		\vertex [right=of p4] (p5) {\(q'\)};
		\vertex [below=of p]  (pb) {\(-q'\)};
		\vertex [below=of p1] (p1b);
		\vertex [below=of p2] (p2b);
		\vertex [below=of p3] (p3b);
		\vertex [below=of p4] (p4b);
		\vertex [below=of p5] (p5b) {\(-q\)};
		\diagram* {
			(p) -- [fermion] (p1) -- [fermion] (p2) -- [scalar] (p3) -- [fermion] (p4) -- [fermion] (p5),
			(p1) -- [boson] (p4b),
			(p2) -- [boson] (p3b),
			(p3) -- [boson] (p2b),
			(p4) -- [boson] (p1b),
			(pb) -- [anti fermion] (p1b) -- [anti fermion] (p2b) -- [scalar] (p3b) -- [anti fermion] (p4b) -- [anti fermion] (p5b)
		};
	\end{feynman}
\end{tikzpicture}
\caption{The 'wheel' diagram in particle-hole channel.}
\label{fig:wheel}
\end{figure}

Thus, the 'wheel' diagram represented in Fig.\;\ref{fig:wheel} effectively transforms into the 'ladder' diagram in Fig.\;\ref{fig:ladder},

\begin{figure}
\begin{tikzpicture}
	\begin{feynman}
		\vertex (p) {\(q\)};
		\vertex [right=of p] (p1);
		\vertex [right=of p1] (p2); 
		\vertex [right=of p2] (p3);
		\vertex [right=of p3] (p4);
		\vertex [right=of p4] (p5) {\(q'\)};
		\vertex [below=of p]  (pb) {\(-q\)};
		\vertex [below=of p1] (p1b);
		\vertex [below=of p2] (p2b);
		\vertex [below=of p3] (p3b);
		\vertex [below=of p4] (p4b);
		\vertex [below=of p5] (p5b) {\(-q'\)};
		\diagram* {
			(p) -- [fermion] (p1) -- [fermion] (p2) -- [scalar] (p3) -- [fermion] (p4) -- [fermion] (p5),
			(p1) -- [boson] (p1b),
			(p2) -- [boson] (p2b),
			(p3) -- [boson] (p3b),
			(p4) -- [boson] (p4b),
			(pb) -- [fermion] (p1b) -- [fermion] (p2b) -- [scalar] (p3b) -- [fermion] (p4b) -- [fermion] (p5b)
		};
	\end{feynman}
\end{tikzpicture}
\caption{The 'ladder' diagram in particle-particle channel.}
\label{fig:ladder}
\end{figure}

\begin{align}
\begin{split}
\sum_{\mathbf{q},...,\mathbf{q}_{k+1},\mathbf{q}',...,\mathbf{q}_{k+1}'}\frac{-(2k)!!}{ (N\beta)^{k} }&{\rm Tr}[\mathbf{G}_{\mathbf{q}_{k+1},\mathbf{q}_k}...\mathbf{RG}_{\mathbf{q}_1,\mathbf{q};\mathbf{p}_1,r_1}\mathbf{\Delta}_{\mathbf{q},\mathbf{q}'}\mathbf{RG}^T_{\mathbf{q}',\mathbf{q}_1';\mathbf{p}_1',r_1'}...\mathbf{RG}^T_{\mathbf{q}_{k-1}',\mathbf{q}_{k}';\mathbf{p}',r'}\mathbf{G}^T_{\mathbf{q}_{k}',\mathbf{q}_{k+1}'}\mathbf{\Delta}^{\dag}_{\mathbf{q}_{k+1}',\mathbf{q}_{k+1}}]\times\\
&\times W^{(\mathbf{p}_1,r_1),(\mathbf{p}_1',r_1')}...W^{(\mathbf{p}_k,r_k),(\mathbf{p}_k',r_k')}.
\end{split}
\end{align}
Considering a $\mathbf{\Delta}$ such that 
\begin{equation}
\frac{1}{ N\beta }\sum_{\mathbf{q}_1,\mathbf{q}_1'}\mathbf{RG}_{\mathbf{q}_2,\mathbf{q}_1;\mathbf{p}_2,r_2}\mathbf{\Delta}_{\mathbf{q}_1,\mathbf{q}_1'}\mathbf{RG}^T_{\mathbf{q}_1',\mathbf{q}_2';\mathbf{p}_2',r_2'}W^{(\mathbf{p}_2,r_2),(\mathbf{p}_2',r_2')} = \lambda\mathbf{\Delta}_{\mathbf{q}_2,\mathbf{q}_2'},
\end{equation}
the expression simplifies to
\begin{equation} \label{Cris}
 -(2k)!!\lambda^k\sum_{\mathbf{q}_{k},\mathbf{q}_{k+1},\mathbf{q}_{k}',\mathbf{q}_{k+1}'} {\rm Tr}[\mathbf{G}_{\mathbf{q}_{k+1},\mathbf{q}_k}\mathbf{\Delta}_{\mathbf{q}_{k},\mathbf{q}_{k}'}\mathbf{G}^T_{\mathbf{q}_{k}',\mathbf{q}_{k+1}'}\mathbf{\Delta}^{\dag}_{\mathbf{q}_{k+1}',\mathbf{q}_{k+1}}].
\end{equation}
After substituting Eq.\;(\ref{Cris}) into Eq.\;(\ref{Stefan}) and performing a resummation of the leading terms, we obtain 
\begin{equation}\label{resummation}
{\rm Tr}[\mathbf{W} \delta^2\mathbf{W}^{-1}] = \frac{2\lambda}{1-\lambda}\sum_{\mathbf{q}_{1},\mathbf{q}_{2},\mathbf{q}_{1}',\mathbf{q}_{2}'} {\rm Tr}[\mathbf{G}_{\mathbf{q}_{2},\mathbf{q}_1}\mathbf{\Delta}_{\mathbf{q}_{1},\mathbf{q}_{1}'}\mathbf{G}^T_{\mathbf{q}_{1}',\mathbf{q}_{2}'}\mathbf{\Delta}^{\dag}_{\mathbf{q}_{2}',\mathbf{q}_{2}}]+\cal{O}(\lambda),
\end{equation}
where we used 
\begin{equation}
\sum_{k=1}\lambda^k=\frac{\lambda}{1-\lambda}.
\end{equation}
Eq.\;(\ref{resummation}) implies that the instability appears when $\lambda = 1$. Therefore, the instability condition is
\begin{equation}\label{gap_generic}
\frac{1}{ N\beta}\sum_{\mathbf{q}_1,\mathbf{q}_1'}\mathbf{R}_{\mathbf{q}_2,\mathbf{q}_2-\mathbf{p};\mathbf{p},r}\mathbf{G}_{\mathbf{q}_2-\mathbf{p},\mathbf{q}_1}\mathbf{\Delta}_{\mathbf{q}_1,\mathbf{q}_1'}\mathbf{G}^T_{\mathbf{q}_1',\mathbf{q}_2'-\mathbf{p}'}\mathbf{R}^T_{\mathbf{q}_2'-\mathbf{p}',\mathbf{q}_2';\mathbf{p}',r'}W^{(\mathbf{p},r),(\mathbf{p}',r')} = \mathbf{\Delta}_{\mathbf{q}_2,\mathbf{q}_2'},
\end{equation}
for some $\mathbf{\Delta}$. The condition can be graphically represented by Fig.\;\ref{fig:BetheS_ladder} and corresponds to the generalized gap equation (implying the Cooper instability condition), in the limit of an infinitesimal gap or, equivalently, to the Bethe-Salpeter equation\cite{Nambu,Salpeter_Bethe} in the 'ladder' approximation with renormalized interaction $\mathbf{W}$ and propagator $\mathbf{G}$, evaluated in the presence of the effective field $\mathbf{M}_{\rm sp}$.

\begin{figure}
	\begin{tikzpicture}
		\begin{feynman}
			\vertex (p); 
			\vertex [right=of p] (p0);
			\vertex [right=of p0] (p1);
			\vertex [right=of p1] (p2); 
			\vertex [right=of p2] (p3); 
			\vertex [below=of p]  (pb);
			\vertex [below=of p0] (p0b);
			\vertex [below=of p1] (p1b); 
			\vertex [below=of p2] (p2b);
			\vertex [below=of p3] (p3b);
			\diagram* {
				(p) -- (p0),
				(p0) -- [scalar, edge label={\qquad \(=\)}] (p0b),
				(pb) -- (p0b),
				(p1) -- (p2),
				(p1b) -- (p2b),
				(p2) -- [fermion] (p3),
				(p2) -- [boson] (p2b),
				(p2b) -- [fermion] (p3b),
				(p3) -- [scalar] (p3b)
			};
		\end{feynman}
	\end{tikzpicture}
	\caption{The Bethe-Salpeter equation in the 'ladder' approximation.}
	\label{fig:BetheS_ladder}
\end{figure}

\begin{figure}
	\begin{subfigure}[b]{0.45\textwidth}
		\centering 
		\begin{tikzpicture}
			\begin{feynman}
				\vertex (p1);
				\vertex [right=of p1] (p2); 
				\vertex [right=of p2] (p3);
				\vertex [right=of p3] (p4);
				\vertex [right=of p4] (p5);
				\vertex [below=of p1] (p1b);
				\vertex [below=of p2] (p2b);
				\vertex [below=of p3] (p3b);
				\vertex [below=of p4] (p4b);
				\vertex [below=of p5] (p5b);
				\diagram* {
					(p1) -- (p2) -- [fermion] (p3) -- [fermion] (p4) -- [fermion] (p5),
					(p2) -- [boson] (p2b),
					(p3) -- [boson, half left] (p4),
					(p1b) -- (p2b) -- [fermion] (p5b),
					(p5) -- [scalar] (p5b)
				};
			\end{feynman}
		\end{tikzpicture}
		\caption{}
		\label{fig:fermion_propagator_renormalization}
	\end{subfigure}
	\hfill
	\begin{subfigure}[b]{0.45\textwidth}
		\centering 
		\begin{tikzpicture}
			\begin{feynman}
				\vertex (p);
				\vertex [right=of p] (p2); 
				\vertex [right=of p2] (p3);
				\vertex [right=of p3] (p4);
				\vertex [right=of p4] (p5);
				\vertex [below=of p] (pb);
				\vertex [below=of p2] (p2b);
				\vertex [below=of p3] (p3b);
				\vertex [below=of p4] (p4b);
				\vertex [below=of p5] (p5b);
				\diagram* {
					(p) -- (p2) -- [fermion] (p3) -- [fermion] (p4) -- [fermion] (p5),
					(p3) -- [boson] (p3b),
					(p2) -- [boson, half left] (p4),
					(pb) -- (p3b) -- [fermion] (p5b),
					(p5) -- [scalar] (p5b)
				};
			\end{feynman}
		\end{tikzpicture}
		\caption{}
		\label{fig:vertex_renormalization}
	\end{subfigure}
	\caption{Fermion propagator renormalization (a) and vertex renormalization (b).}
	\label{fig:propagator_vertex_renormalization}
\end{figure}


\begin{figure}
	\begin{tikzpicture}
		\begin{feynman}
			\vertex (p1);
			\vertex [right=of p1] (p2); 
			\vertex [right=of p2] (p3);
			\vertex [right=of p3] (p4);
			\vertex [below=of p1] (p1b);
			\vertex [below=of p2] (p2b);
			\vertex [below=of p3] (p3b);
			\vertex [below=of p4] (p4b);
			\diagram* {
				(p1) -- (p2) -- [fermion] (p3) -- [fermion] (p4),
				(p2) -- [boson] (p3b),
				(p3) -- [boson] (p2b),
				(p1b) -- (p2b) -- [fermion] (p3b) -- [fermion] (p4b),
				(p4) -- [scalar] (p4b)
			};
		\end{feynman}
	\end{tikzpicture}
	\caption{Cross term.}
	\label{fig:BetheS_ladder_cross}
\end{figure}
Taking into account the other terms in the sum in Eq.\;(\ref{Stefan}) implies that all two particle irreducible diagrams should be also included in the instability condition given by Eq.\;(\ref{gap_generic}) (without fermionic loops in the leading order approximation), such as Fig.\;\ref{fig:fermion_propagator_renormalization}, Fig.\;\ref{fig:vertex_renormalization}, and Fig.\;\ref{fig:BetheS_ladder_cross}. The first two diagrams correspond to the lowest order contribution to the renormalized fermion propagator, denoted $\bm{G}$, and the interaction vertex, denoted $\bm{R}$, correspondingly. According to the Dyson equation $\bm{G}^{-1} = \mathbf{G}^{-1} - \mathbf{\Sigma}$, where the self-energy $\mathbf{\Sigma}$ (exchange part) is calculated with the screened interaction $\mathbf{W}$ and provides additional contribution in $\bm{G}$, besides the effective field $\mathbf{M}_{\rm sp}$. The self-consistent equation in the lowest order 
\begin{equation}\label{Sigma_GW}
	\mathbf{\Sigma}_{\mathbf{q},\mathbf{q}'} = \sum_{\mathbf{q}_1,\mathbf{q}_2}\bm{R}_{\mathbf{q},\mathbf{q}_1;\mathbf{p},r}\bm{G}_{\mathbf{q}_1,\mathbf{q}_2}\bm{R}_{\mathbf{q}_2,\mathbf{q}';\mathbf{p}',r'}W^{(\mathbf{p},r),(\mathbf{p}',r')},
\end{equation}
is analogous to the $GW$ approximation, \cite{Hedin_1965, Hedin_Lundqvist_1971} which is sufficient for an approximate effective mass calculation in heavy fermion systems.\cite{Zhang_DasSarma_2005} A peculiarity of this approximation is that the local part ("Hartree part") is calculated with the bare interaction $\mathbf{U}$, while the correlation part ("Fock part") is calculated with the screened interaction $\mathbf{W}$. Higher order terms can be written down in a similar manner. The renormalized self-consistent vertex $\bm{R}$ in the lowest order reads 
\begin{equation}\label{vertex_R}
	\bm{R}_{\mathbf{q},\mathbf{q}';\mathbf{p},r} = \mathbf{R}_{\mathbf{q},\mathbf{q}';\mathbf{p},r}+ \sum_{\mathbf{q}_1,...,\mathbf{q}_4}\bm{R}_{\mathbf{q},\mathbf{q}_1;\mathbf{p}',r'}\bm{G}_{\mathbf{q}_1,\mathbf{q}_2}\bm{R}_{\mathbf{q}_2,\mathbf{q}_3;\mathbf{p},r}\bm{G}_{\mathbf{q}_3,\mathbf{q}_4}\bm{R}_{\mathbf{q}_4,\mathbf{q}';\mathbf{p}'',r''}W^{(\mathbf{p}',r'),(\mathbf{p}'',r'')}.
\end{equation}
A similar approximation was used in [\onlinecite{Onari_Kontani}]. The use of the term 'renormalized' for the propagator, interaction, and vertex is justified also in the context of quantum electrodynamics. This is because they are directly related to measurable quantities, such as the physical mass, electron charge, and spin, correspondingly, while the divergent parts in Eqs.\;(\ref{W_exact}), (\ref{Sigma_GW}), and (\ref{vertex_R}) are absorbed into redefinition of the bare  $\mathbf{G}$, $\mathbf{U}$, and $\mathbf{R}$. In this sense, the renormalization procedure can be interpreted as a perturbative solution of the system of self-consistent equations with redefinition of the bare values. Taking into account the propagator and vertex renormalization, the Bethe-Salpeter equation becomes
\begin{equation}\label{gap_generic_renorm}
	\frac{1}{ N\beta}\sum_{\mathbf{q},\mathbf{q}_1,\mathbf{q}',\mathbf{q}_1'}\bm{R}_{\mathbf{q}_2,\mathbf{q}_1;\mathbf{p},r}\bm{G}_{\mathbf{q}_1,\mathbf{q}}\mathbf{\Delta}_{\mathbf{q},\mathbf{q}'}\bm{G}^T_{\mathbf{q}',\mathbf{q}_1'}\bm{R}^T_{\mathbf{q}_1',\mathbf{q}_2';\mathbf{p}',r'}W^{(\mathbf{p},r),(\mathbf{p}',r')} = \mathbf{\Delta}_{\mathbf{q}_2,\mathbf{q}_2'}.
\end{equation}
This equation has a rather generic form, and does not rely on specific assumptions about the structure of $\bm{G}$, such as finite effective mass. However, Eq.\;(\ref{gap_generic_renorm}) does not include the contribution from the crossed terms, mentioned above (Fig.\;\ref{fig:BetheS_ladder_cross}):
\begin{equation}\label{gap_generic_renorm_cross}
	\frac{1}{ N\beta}\sum_{\mathbf{q},\mathbf{q}_1,\mathbf{q}',\mathbf{q}_1'}\bm{RG}_{\mathbf{q}_2,\mathbf{q}_1;\mathbf{p}_2,r_2}\bm{RG}_{\mathbf{q}_1,\mathbf{q};\mathbf{p}_1,r_1}\mathbf{\Delta}_{\mathbf{q},\mathbf{q}'}\bm{RG}^T_{\mathbf{q}',\mathbf{q}_1';\mathbf{p}_1',r_1'}\bm{RG}^T_{\mathbf{q}_1',\mathbf{q}_2';\mathbf{p}_2',r_2'}W^{(\mathbf{p}_1,r_1),(\mathbf{p}_2',r_2')}W^{(\mathbf{p}_2,r_2),(\mathbf{p}_1',r_1')}.
\end{equation}
As it was shown in Refs.\ [\onlinecite{Berk_Schrieffer,Doniach_Engelsberg}], it is possible to perform a re-summation of the cross terms (as in Fig.\;\ref{fig:BetheS_ladder_cross}) for some specific simple cases, i.e. with on-site interaction and commuting Hamiltonian, which leads to a renormalization of the interaction and reflects the exchange symmetry. However, focusing mainly on the properties of the effective interaction $\mathbf{W}$ beyond RPA, below we omit the cross terms from further consideration. In particular, for a homogeneous system with $\bm{G}_{\mathbf{q},\mathbf{q}'} = \bm{G}(\mathbf{q})\delta_{\mathbf{q},\mathbf{q}'}$, $W^{(\mathbf{p},r),(\mathbf{p}',r')} = W^{r r'}(\mathbf{p})\delta_{\mathbf{p},-\mathbf{p}'}$ and for $\mathbf{\Delta}_{\mathbf{q}_1,\mathbf{q}_1'}=\mathbf{\Delta}(\mathbf{q}_1)\delta_{\mathbf{q}_1,-\mathbf{q}_1'}$, neglecting vertex renormalization and substituting Eq.\;(\ref{P_matrices}), the equation substantially simplifies,
\begin{equation}
\frac{1}{ N\hbar^2\beta}\sum_{\mathbf{q}_1,r,r'}\mathbf{P}^{r} \eta_{r r} \bm{G}(\mathbf{q}_1)\mathbf{\Delta}(\mathbf{q}_1)\bm{G}^T(-\mathbf{q}_1)(\mathbf{P}^{r'})^T \eta_{r' r'}W^{r r'}(\mathbf{q}_2-\mathbf{q}_1) = \mathbf{\Delta}(\mathbf{q}_2),
\end{equation}
where we substitute $\mathbf{p} = \mathbf{q}_2 - \mathbf{q}_1$. Furthermore, let us assume that $\bm{G}$ and $\mathbf{P}$ commute. In this case
\begin{equation}\label{gap_commutative}
	\frac{1}{ N\hbar^2\beta}\sum_{\mathbf{q}_1,r,r'}\bm{G}(\mathbf{q}_1)\mathbf{P}^{r}  \mathbf{\Delta}(\mathbf{q}_1)(\mathbf{P}^{r'})^T \bm{G}^T(-\mathbf{q}_1)\eta_{r r}\eta_{r' r'}W^{r r'}(\mathbf{q}_2-\mathbf{q}_1) = \mathbf{\Delta}(\mathbf{q}_2).
\end{equation}
\end{widetext}
Now, we consider singlet pairing. The property of Pauli matrices,
\begin{equation}
	\sigma_{i}^T\sigma_{2}\sigma_{j} +\sigma_{j}^T\sigma_{2}\sigma_{i}  = -2\sigma_{2}\delta_{ij},
\end{equation}
leads to
\begin{equation}
	(\mathbf{P}^{r})^T \mathbf{P}^{3} \mathbf{P}^{r'} +(\mathbf{P}^{r'})^T \mathbf{P}^{3} \mathbf{P}^{r} = -2\mathbf{P}^{3}\eta_{r r'} ,
\end{equation}
for $r,r' \in A$, with an analogous expression for $\mathbf{P}^{7}$ for $r,r' \in B$. Therefore, we look for a spin-dependent solution in the form (singlet pairing,  with $\mathbf{\Delta} \sim \sigma_{2}$)
\begin{equation}
	\mathbf{\Delta}(\mathbf{q}) = (\mathbf{P}^{3}+\mathbf{P}^{7})\Delta(\mathbf{q}).
\end{equation}
In this case, introducing the average interaction
\begin{equation}\label{W_avg}
	W(\mathbf{p})=	\frac{1}{ 2}\sum_{r,r' } \eta_{r' r}W^{r r'}(\mathbf{p}),
\end{equation}
the equation simplifies further
\begin{equation}\label{gap_homog}
	-\frac{1}{ N\hbar^2\beta}\sum_{\mathbf{q}_1}W(\mathbf{q}_2-\mathbf{q}_1)\bm{G}(\mathbf{q}_1) \mathbf{\Delta}(\mathbf{q}_1) \bm{G}^T(-\mathbf{q}_1) = \mathbf{\Delta}(\mathbf{q}_2).
\end{equation}
The instability condition and, therefore, the average interaction $W(\mathbf{p})$ should be  independent from the choice of $\mathbf{m}$. However, this is true only in the lowest order of $U$. Therefore, for a rotation invariant system, we set $\mathbf{m}$ to a particular value that corresponds to the Hubbard-Stratonovich transformation of the form $c^{\dag}_{\mathbf{j},\uparrow}c^{\dag}_{\mathbf{j},\downarrow}c_{\mathbf{j},\downarrow}c_{\mathbf{j},\uparrow}=
n_\mathbf{j}^2/4-S_{\mathbf{j},z}^2$.

The original result of Refs.\ [\onlinecite{Berk_Schrieffer}, \onlinecite{Doniach_Engelsberg}] corresponds to the case when $W(\mathbf{p})$ is approximated at RPA level in Eq.\;(\ref{gap_homog}), yielding
\begin{equation}\label{W_avg_RPA}
	W(\mathbf{p})=	\frac{U}{ 1-U^2\chi^2(\mathbf{p})},
\end{equation} 
where $\chi(\mathbf{p})$ is the Lindhard particle-hole susceptibility. Moreover, the bare interaction $\mathbf{U}$ is assumed for the cross terms, which makes it possible to perform their re-summation analytically for the particle-particle vertex
\begin{equation}\label{V_with_cross}
V_{\mathbf{k},\mathbf{k'}}	= W(\mathbf{k}-\mathbf{k'})+\frac{U^2\chi(\mathbf{\mathbf{k}+\mathbf{k'}})}{ 1-U\chi(\mathbf{\mathbf{k}+\mathbf{k'}})}.
\end{equation}
Clearly, this approach is somewhat inconsistent in treating the interaction for the cross terms, where the effective interaction $W$ should be used instead of $U$.

Now we consider two special cases:\\
1) When there is neither interaction nor hopping between identical sublattices, the corresponding equations decouple,

\begin{equation}\label{gap_homog_scalar}
-\frac{1}{ N\hbar^2\beta}\sum_{\mathbf{q}_1}W(\mathbf{q}_2-\mathbf{q}_1)G(\mathbf{q}_1)G(-\mathbf{q}_1) \Delta(\mathbf{q}_1)  = \Delta(\mathbf{q}_2).
\end{equation}
In case $W(\mathbf{q})$ is a slowly varying function, there exists a solution $\Delta(\mathbf{q})$ to Eq.\;(\ref{gap_homog_scalar}) that is also a slowly varying function, if the following condition holds
\begin{equation}
 -\frac{1}{ N\hbar^2\beta }\sum_{\mathbf{q}}G(\mathbf{q})G(-\mathbf{q})W(\mathbf{q}) = 1.
\end{equation}
This is equivalent to the Cooper instability condition, establishing a relation between the critical temperature $T_c$ and the frequency $\omega_{D}$ of the renormalized interaction $W$ (Debye frequency),
\begin{equation}
k_{B}T_c = \frac{2e^{\gamma}}{\pi}\omega_{D}\exp\left[-\frac{1}{W_0 D(0)}\right],
\end{equation}
where $\gamma$ is the Euler constant, $W_0 = W(0)$ , and $D(0)$ is the density of states at the Fermi energy. Above the Debye frequency, $W$ vanishes. \\
2) When $W(\mathbf{q})$ does not depend on the Matsubara frequency, i.e. $W(\mathbf{q})=W_{\mathbf{k}}$. In this case, $\Delta(\mathbf{q})=\Delta_{\mathbf{k}}$ does not depend on the frequency either, and we perform a  Matsubara summation in Eq.\;(\ref{gap_homog_scalar}) following the same procedure as described in the Appendix \ref{appendix}. Taking into account that $\mathbf{H}_{\mathbf{-k}}=\mathbf{H}^T_{\mathbf{k}}$, the summation yields
\begin{equation}\label{gap_Fermi}
-\frac{1}{ N}\sum_{\mathbf{k}'}W_{\mathbf{k}-\mathbf{k}'} \frac{1-2n_F(E_{\mathbf{k}'})}{2E_{\mathbf{k}'}}\Delta_{\mathbf{k}'} = \Delta_\mathbf{k},
\end{equation}
which corresponds to the BCS gap equation in the limit of an infinitesimal gap, with a difference that, Eq.\;(\ref{gap_Fermi}) involves the poles $E_{\mathbf{k}}$ of the renormalized propagator $\bm{G}$ and the renormalized interaction $W$ instead of the bare $U$ that usually enters mean-filed theory results. It is particularly interesting to analyze Eq.\;(\ref{gap_Fermi}) for solutions with nontrivial dependence on momentum, in particular, of the form,
\begin{equation}
 \sum_{E_{\mathbf{k}}=E}\Delta_\mathbf{k} = 0,
\end{equation}
for any energy level $E$. In this case, any shift of $W$ is irrelevant,\cite{Timm}
\begin{equation}\label{gap_zero}
\sum_{\mathbf{k}'}W_{0} \frac{1-2n_F(E_{\mathbf{k}'})}{2E_{\mathbf{k}'}}\Delta_{\mathbf{k}'} = 0.
\end{equation}

In Eq.\;(\ref{gap_Fermi}), $W_{\mathbf{k}}$ is a renormalized interaction, which accounts for screening. Namely, for square a lattice close to half-filling, it is weakly repulsive for small momenta, but strongly repulsive for the values leading to nesting of the Fermi surface that leads to a $d$-wave gap solution.\cite{Cyrot, Scalapino_Loh_Hirsch_1986, Miyake_1986} If present, a flat band close to the Fermi surface has a twofold impact on the gap equation. Firstly, it decreases the denominator, reducing the kinetic energy cost of superconducting gap formation.\cite{Cao1, Cao2, GuohongLi, Pixley,Nunes_MoraisSmith} Secondly, it increases the anisotropy of the effective interaction,\cite{Makogon_2011,Pizarro,Goodwin} which is also favorable for the unconventional pairing, 
as suggested in Ref.\;[\onlinecite{Gonzalez}] and in Ref.\;[\onlinecite{Samajdar_Scheurer}], according to a weak-coupling approximation. The results presented here confirm that this twofold favorable impact persists beyond the weak-coupling approximation.

\section{Conclusions}
We developed a novel median-point approximation method for evaluating the partition function, thus generalizing the commonly used Laplace's method (saddle-point approximation), which approximates the given distribution function by a Gaussian  around the mode. Our method provides a different perspective, considering a manifold of distributions diffeomorphic to a Gaussian distribution. Clearly, the mode of the Gaussian distribution coincides with the median point and remains median in any of the diffeomorphic distributions. Matching, up to a certain order, the derivatives of the logarithm of the density function diffeomorphic to a Gaussian distribution and the given one at the median point provides a generalization to the Laplace's method, where only the first two derivatives are matched. In the case of one-dimensional distributions, the matching conditions form a system of nonlinear equations that can be closed at a certain order, as an approximation, with some constraints on higher derivatives of the mapping function. In case the median point is known, the system reduces to a polynomial equation for the precision of the Gaussian distribution (the inverse of the variance). We demonstrate that the largest real root of the polynomial equation converges to the exact value with increasing order of approximation. An interesting open question remains if this is always the case. 

Considering a system of fermions with local interactions on a lattice and performing a Hubbard-Stratonovich transformation, we applied the method for the evaluation of the corresponding bosonic partition function, after integrating out the fermionic fields. Matching of all derivatives yields an infinite system of nonlinear equations, describing the mapping to the corresponding quadratic bosonic action, i.e. a free boson distribution. The distribution is characterized by the median point, interpreted as an effective field at which the fermionic propagators are evaluated, and the bosonic propagator is interpreted as the effective interaction. The corresponding exact expressions for the effective field and interactions have the form of the saddle-point approximation values, with additional terms originating from the Jacobian of the transformation. In particular, we identify the terms giving rise to Maki-Thompson and Aslamazov-Larkin type corrections to the effective interaction and  demonstrate that the linear relationship between the effective interaction and susceptibility is an artifact of RPA, which does not hold in general, when higher order interaction terms are taken into account and that the two quantities do not necessarily diverge simultaneously. Moreover, we demonstrate that the effective interaction should remain finite even at a phase transition point. A solution of the system of equations up to the leading-order approximation corresponds to a sum of diagrams with one fermionic loop. Within the leading-order approximation, the response of the system to the presence of an infinitesimally small pairing interaction involves the resummation of an infinite number of diagrams with 'wheel' structure in the particle-hole channel. By establishing the equivalence of the 'wheel' diagrams in the particle-hole channel to the 'ladder' diagrams in the particle-particle channel, the resummation of these leads to the Bethe-Salpeter equation, or equivalently, to the generalized gap equation in the limit of an infinitesimal gap (implying the Cooper instability condition). Thus, we demonstrate the emergence of a pairing instability as an interaction instability in the particle-hole channel, without explicitly introducing the particle-particle channel and show that the interaction enters the Bethe-Salpeter equation only in its effective form to all orders, including the exchange part of the self-energy (contrary to the typical mean-field results). Besides the renormalized propagator and interaction, the vertex function is also renormalized. We demonstrate that solving perturbatively the system of self-consistent equations for the three quantities in quantum electrodynamics is analogous to the renormalization procedure (at least, at one loop level), justifying the use of the term "renormalized". Analyzing the Bethe-Salpeter equation, we find that the presence of a flat band close the Fermi level has a twofold favorable impact, persisting beyond the weak-coupling approximation: a reduced kinetic energy cost of gap formation and an increased anisotropy of the effective interaction, favoring a momentum dependent order parameter. 

Summarizing, within the median-point approximation framework, CDW and SDW correspond to a shift of the median point, while the superconducting phase transition corresponds to an instability of the renormalized interaction towards the appearance of pairing interaction (that depends on the inbound momentum). 
 
\section{Acknowledgments}
The authors have greatly benefited from fruitful and inspiring discussions with Vladimir Gritsev. Moreover, we are grateful to the contributors of the SymPy computer algebra system that was very helpful for deriving the scalar function results. In addition, we are thankful to the creator of the package we used for drawing Feynman diagrams.\cite{JoshuaEllis} D.M. acknowledges Utrecht University for providing the possibility to work on this subject during 2010-2011, when most of the presented results were obtained.

\appendix
\section{Median-point approximation for one-dimensional integrals}
\label{one_dimensional_integrals}
Until now, we used the multivariate approximation to the lowest order beyond the saddle-point approximation (one fermionic loop). Now, we will take into account higher orders, but specifically for one-dimensional integrals for simplicity. 

For convenience, let us consider the function
\begin{equation}
	F(x_0) = \int_{-\infty}^{x_0}\exp[-f(x)]\ud x,
\end{equation}
where $f(x) \in R$ is a real function and, clearly, $f(x) = - \ln [-F'(x)]$. We are interested to find the value of the integral when $x_0 \rightarrow \infty$, i.e. the value of $F(\infty)$ (clearly, $F(-\infty) = 0$). In general, the function $F(x)$ can not be expressed in terms of elementary functions and some approximations have to be made. One of the simplest approximations used in such cases is the Laplace's approximation. However, the approximation is not effective or not applicable if the second derivative vanishes (or is close to zero) at the critical point $x_c$ (defined by $f'(x_c)=0$). Therefore, we consider a generalization of the Laplace's method. First, we consider an invertible change of the integration variable from $x$ to $y$ related as $y(x)$ (and having an inverse function $x(y)$ with $x'(0)=1$), with $y(\pm\infty) = \pm\infty$. Then, we denote $y(x_0) = y_0$ and the integral becomes
\begin{equation}
	F(x_0) = \int_{-\infty}^{y(x_0)}\exp[-f\textbf{(}x(y)\textbf{)}+\ln\textbf{(}x'(y)\textbf{)}]\ud y.
\end{equation}
For convenience, we denote $g(y) = f\textbf{(}x(y)\textbf{)}-\ln\textbf{(}x'(y)\textbf{)}$ such that
\begin{equation}
	F(x_0) = \int_{-\infty}^{y_0}\exp[-g(y)]\ud y
\end{equation}
Suppose that the change of the integration variable can be done, such that the function $g(y)$ has the form
\begin{equation}
	g(y) = g(0) + \frac{\gamma}{2} y^2 +r_n(y),
\end{equation}
where derivatives $r^{(k)}_n(0) =0$ for $k\leq n$ ($r_n(0) = 0$ by definition) for some $n$. This implies the relation
\begin{equation}
	f\textbf{(}x(y)\textbf{)}-\ln\textbf{(}x'(y)\textbf{)} = f\textbf{(}x(0)\textbf{)} + \frac{\gamma}{2} y^2 +r_n(y).
\end{equation}
Alternatively, we can consider the variable transformation not as $x(y)$ but as $y(x)$, which often turns out to be more convenient. Denoting $x_m = x(0)$, the relation reads
\begin{equation}\label{gamma_y2_x}
	\frac{\gamma}{2} y^2(x) = f(x)+\ln\textbf{(}y'(x)\textbf{)}  - f(x_m) - \tilde{r}_n(x),
\end{equation}
where $\tilde{r}_n(x)\equiv r_n\textbf{(}y(x)\textbf{)}$. One observes that $\tilde{r}^{(k)}_n(x_m) =0$ for $k\leq n$ (with differentiation now performed with respect to $x$). This allows us to give an alternative interpretation to the same change of variables. Namely, the change of variables in the Gaussian integral reads
\begin{equation}
	\int_{-\infty}^{y_0}\exp\left(-\frac{\gamma}{2} y^2\right)\ud y =  \int_{-\infty}^{x_0}\exp[-f(x)+ f(x_m) + \tilde{r}_n(x)]\ud x.
\end{equation}
Defining
\begin{equation}\label{F_n_def}
	F_n(x_0) \equiv \int_{-\infty}^{x_0}\exp[-f(x)+ \tilde{r}_n(x)]\ud x
\end{equation}
and recalling that
\begin{equation}
	\int_{-\infty}^{y_0}\exp\left(-\frac{\gamma}{2} y^2\right)\ud y = \sqrt{\frac{2\pi}{\gamma}}\Phi(\sqrt{\gamma}y_0),
\end{equation}
where $\Phi(x)$ is the cumulative distribution function of the standard normal distribution, we can give a formal explicit relation for the variable change, having
\begin{equation}\label{phi_y_f_x}
	\sqrt{\frac{2\pi}{\gamma}}\Phi(\sqrt{\gamma}y) = F_n(x) \exp[f(x_m)].
\end{equation}
Evaluating Eq.\;(\ref{phi_y_f_x}) at $y=0$ results into
\begin{equation}\label{F_x_m}
	\sqrt{\frac{\pi}{2\gamma}} = F_n(x_m) \exp[f(x_m)],
\end{equation}
whereas evaluating at $y\rightarrow \infty$ leads to
\begin{equation}\label{F_infty}
	\sqrt{\frac{2\pi}{\gamma}} = F_n(\infty) \exp[f(x_m)].
\end{equation}
Combining Eq.\;(\ref{F_x_m}) and Eq.\;(\ref{F_infty}) yields
\begin{equation}
	F_n(\infty) = 2 F_n(x_m),
\end{equation}
which confirms that $x_m$ is indeed the median point for $F_n(x)$, as expected for the invertible change of variables.
Substituting Eq.\;(\ref{F_x_m}) into Eq.\;(\ref{phi_y_f_x}) and applying the inverse of the cumulative distribution function yields
\begin{equation}
	y = \frac{1}{\sqrt{\gamma}}\Phi^{-1}\left(\frac{F_n(x)}{F_n(\infty)}\right).
\end{equation}
The value of $\gamma$ is such that $y'(x_m)=1$. It follows from Eq.\;(\ref{F_infty}) that
\begin{equation}
	F_n(\infty) = \sqrt{\frac{2\pi}{\gamma}} \exp[-f(x_m)]
\end{equation}
and to find the value $F_n(\infty)$ it is sufficient to know $\gamma$ and $x_m$. Clearly, for the case $n\rightarrow \infty$, i.e. $\tilde{r}(x) =0$, one sees in Eq.\;(\ref{F_n_def}) that
\begin{equation}
	F_n(x) \rightarrow  F(x). 
\end{equation}
In this limit, the expression for the new variables is not very helpful when $F(x)$ is unknown and the objective is to find $F(\infty)$. However, we can consider the variable transformation for a certain $n$ and use it to find an approximation for $F(\infty)$ that would approach the exact value in the limit $n\rightarrow \infty$. We can achieve this by differentiating Eq.\;(\ref{gamma_y2_x}) with respect to $x$ and evaluating it at the point $x_m$. For illustration, the case with $n\geq 4$ is considered. The first derivative is
\begin{equation}
	\gamma y(x)y'(x) = f'(x)+y'(x)^{-1}y''(x)- {\tilde{r}'}_n(x),
\end{equation}
which yields
\begin{equation}
	f'(x_m) = -y''(x_m). 
\end{equation}
\begin{widetext}
	The second derivative is
	\begin{equation}
		\gamma [y(x)y''(x) + y'^2(x)] = f''(x)+y'(x)^{-1}y'''(x) - y'(x)^{-2}y''^2(x) - \tilde{r}''_n(x),
	\end{equation}
	which yields
	\begin{equation}
		\gamma = f''(x_m) + y'''(x_m) - y''^2(x_m). 
	\end{equation}
	Furthermore, the third derivative is
	\begin{equation}
		\gamma [y(x)y'''(x) +
		3y'(x)y''(x)] = f'''(x)+\frac{\ud^3}{{\ud x}^3}\ln\textbf{(}y'(x)\textbf{)} - \tilde{r}'''_n(x),
	\end{equation}
	where
	\begin{equation}
		\frac{\ud^3}{{\ud x}^3}\ln\textbf{(}y'(x)\textbf{)} = y'(x)^{-1}y^{(4)}(x) - 3y'(x)^{-2}y''(x)y'''(x) + 2y'(x)^{-3}y''^3(x),
	\end{equation}
	which yields
	\begin{equation}
		3\gamma y''(x_m) = f'''(x_m) + y^{(4)}(x_m) - 3y''(x_m)y'''(x_m) + 2y''^3(x_m).
	\end{equation}	
\end{widetext}
It follows directly from the equations that in the asymptotic limit $\gamma \rightarrow \infty$ 
\begin{equation}
	y^{(k)}(x_m) \rightarrow  \frac{f^{(k+1)}(x_m)}{(k+1)\gamma}.
\end{equation}
As it was already observed earlier, in order to quantify $F(\infty)$ it is sufficient to know $\gamma$ and $x_m$. These can be found when $y''(x_m)$ and $y'''(x_m)$ are known, which, in turn, depend on $y^{(4)}(x_m)$ and $y^{(5)}(x_m)$, and so on. In other words, the above four equations or any other finite number of equations do not form a closed system. However, in case some finite order $n\geq 3$ is chosen, the first $n$ equations do form an undetermined system of equations with $n+2$ unknown variables. Thus, two additional constrains have to be chosen in order to find a closed solution. $\gamma$ corresponds to a real and non-negative solution. Preferably, the constraints are chosen such that the resulting approximation for $F(\infty)$ would be close to and approach the exact value in the limit $n\rightarrow \infty$. One of the simplest assumptions is to choose $y^{(n)}(x_m) = 0$ and $y^{(n+1)}(x_m) = 0$, which is motivated if the function behavior in the vicinity of $x_m$ is expected to be more important than away from it.

Let us consider a simple example for illustration, with a symmetric function $f(-x) = f(x)$. In this case, the median $x_m = 0$ and $y(-x) = -y(x)$, which implies $y^{(2k)}(0) = 0$ for any positive integer $k$, which yields the two equations (all derivatives are evaluated at zero):
\begin{equation}\label{gamma_f2_y3}
	\gamma = f'' + y''' 
\end{equation}
and
\begin{equation}\label{y3_f4}
	4\gamma y''' = f^{(4)} + y^{(5)}  - 3y'''^2.
\end{equation}
Clearly, in the asymptotic limit $\gamma \rightarrow \infty$
\begin{equation}
	y''' \rightarrow  \frac{f^{(4)}}{4\gamma}.
\end{equation}
Combining Eq.\;(\ref{gamma_f2_y3}) and Eq.\;(\ref{y3_f4}) yields
\begin{equation}
	y^{(5)}= 7\gamma^2 - 10 \gamma f'' + 3 f''^2 - f^{(4)},
\end{equation}
which implies that when $\gamma \rightarrow 0$ following the asymptotic $\gamma \rightarrow \gamma_a$ then,
\begin{equation}
	y^{(5)} \rightarrow  7\gamma_a^2 - 10 \gamma_a f'' + 3 f''^2 - f^{(4)}.
\end{equation}
Moreover, the next order equation is
\begin{equation}
	6 \gamma  y^{(5)}+10\gamma y'''^2 = f^{(6)} + y^{(7)} - 15 y^{(5)}y''' + 30 y'''^3,
\end{equation}
which implies
\begin{equation}
	\gamma y^{(5)} \rightarrow  \frac{f^{(6)}}{6}.
\end{equation}
Considering the case $n = 4$ and choosing the simple constraint $y^{(5)} = 0$ yields the following equation for $\gamma$:
\begin{equation}
	7\gamma^2 - 10 \gamma f'' + 3 f''^2 - f^{(4)} = 0.
\end{equation}
The quadratic equation has two roots. If real, the largest root of the equation is 
\begin{equation}
	\gamma = \frac{1}{7}(5 f'' + \sqrt{ 4 f''^2 +7 f^{(4)}}),
\end{equation}
which is the only solution consistent with the trivial solution $\gamma = f''$ when $f^{(4)} = 0$. Moreover, given the other parameters are kept fixed, in the asymptotic limit $f''\rightarrow \infty$ 
\begin{equation}
	\gamma = f'' + \frac{f^{(4)}}{4 f''} + o\left(\frac{f^{(4)}}{f''}\right),
\end{equation}
i.e. $\gamma \rightarrow f''$ as expected and in line with the asymptotic. However, the approximation breaks at $\alpha=-\sqrt{\frac{f^{(4)}}{3}}$, whereas the Laplace's approximation breaks already at the critical point $f''=0$. In this case, 
\begin{equation}
	\gamma =  \sqrt{ \frac{f^{(4)}}{7}}.
\end{equation}
This result implicitly requires $f^{(4)}\geq 0$, since in the opposite case the convergence of the integral would be driven by even higher derivatives, such as $f^{(6)}$, which would need to be taken into account. 

Assuming the largest real root to be the approximate solution in general, we can check the convergence of the approximation upon increasing its order. One of the simplest cases is 
\begin{equation}
	f(x) = \frac{1}{2}\alpha x^2 + \frac{1}{24}x^4.
\end{equation}
Then, $ f'' = \alpha $
and
$
f^{(4)} = 1
$. 
Clearly, the integral always converges for all values of $\alpha$, which implies that $\gamma > 0$. In the asymptotic limit $\alpha\rightarrow \infty$, 
\begin{equation}
	\gamma \rightarrow \alpha.
\end{equation}
On the other hand, in the asymptotic limit $\alpha\rightarrow -\infty$, according to the Laplace's approximation (and no symmetry breaking),
\begin{equation}
	\gamma \rightarrow - \frac{\alpha}{2}\exp(-3\alpha^2).
\end{equation}  
\begin{widetext}
	When assessing convergence with increasing order at $\alpha \geq0$, the most relevant is the convergence at the critical point $\alpha = 0$, since the approximations converge with growing $\alpha$. The polynomial equations become:
	\begin{align}
		\begin{split}
			&n = 2: \gamma = 0 \\ 
			&n = 4:  -7 \gamma^2 + 1 = 0 \Rightarrow \gamma \approx 0.37796 \\
			&n = 6:  -127\gamma^3 + 21\gamma = 0 \Rightarrow \gamma \approx 0.40664 \\
			&n = 8:  -4369\gamma^4 + 882\gamma^2 - 35 = 0 \Rightarrow \gamma \approx 0.38419 \\
			&n = 10:  -34807\gamma^5 + 8382\gamma^3 - 473\gamma = 0 \Rightarrow \gamma \approx 0.38801 \\
			&n = 12:  -20036983\gamma^6 + 5622903\gamma^4 - 427427\gamma^2 + 5775 = 0 \Rightarrow \gamma \approx 0.39369 \\
			&n = 14:  -2280356863\gamma^7 + 731677947\gamma^5 - 70301231\gamma^3 + 1783551\gamma = 0 \Rightarrow \gamma \approx 0.38856 \\
			&n = 16:  -49020204823\gamma^8 + 17712692972\gamma^6 - 2060481566\gamma^4 + 78545236\gamma^2 - 375375 = 0 \Rightarrow \gamma \approx 0.38909
		\end{split}
	\end{align}
	
\end{widetext}
That indeed demonstrates convergence to the exact value $\beta = \frac{4\pi}{\sqrt{6}}\Gamma(\frac{1}{4})^{-2}$ ($\approx 0.3903$) with increasing the approximation order. Moreover, the parameter range with positive real solutions increases with increasing the order of the approximation, but remains finite. Apparently, this is related to the fact that the exponential asymptotic of $\beta$ for $\alpha\rightarrow -\infty$ can not be captured by a polynomial of a finite degree. 
\section{Evaluation of $\Gamma^{(1)}$ and $\Gamma^{(2)}$}
\label{appendix}
The expressions for $\Gamma^{(1)}$ and $\Gamma^{(2)}$ can be significantly
simplified by performing a Matsubara summation, which is possible in
the basis of eigenfunctions, where the Green's function matrix
becomes diagonal. In particular, for homogeneous systems this basis
is formed by just plane waves, i.e. the Green's function matrix
becomes diagonal in momentum space,
\begin{equation}
\mathbf{G}_{\mathbf{k},n;\mathbf{k}',n'}^{-1}=\mathbf{G}_{\mathbf{k}}^{-1}(i\hbar\omega_n)
\delta_{\mathbf{k},\mathbf{k}'}\delta_{n,n'}.\nonumber
\end{equation}
The Green's function has the structure
\begin{equation}\label{Greenpoles}
\mathbf{G}_{\mathbf{k}}(i\hbar\omega_n)=\sum_{\alpha}\hbar\mathbf{U}_\mathbf{k}
\mathbf{I}_{\alpha}\mathbf{U}_\mathbf{k}^{\dag}(\mu
-\varepsilon_\mathbf{k}^{(\alpha)}+i\hbar\omega_n)^{-1},\nonumber
\end{equation}
where $\mathbf{I}_{\alpha}\equiv {\rm
Diag}(\delta_{\alpha,1},\delta_{\alpha,2},\delta_{\alpha,3},\delta_{\alpha,4})$
and $\mathbf{U}_\mathbf{k}$ is a unitary matrix
$\mathbf{U}_\mathbf{k}=(\mathbf{V}_\mathbf{k}^1,\mathbf{V}_\mathbf{k}^2,\mathbf{V}_\mathbf{k}^3,\mathbf{V}_\mathbf{k}^4)$
composed of the orthonormal eigenvectors of
$\mathbf{G}_{\mathbf{k}}(i\hbar\omega_n)$ normalized to
$|\mathbf{V}_\mathbf{k}^\alpha|=1$, with
$\varepsilon_\mathbf{k}^{(\alpha)}-\mu$ being the poles of the
Green's function. It follows from the definition for $\Gamma^{(1)}$
that
\begin{equation}
\Gamma^{(1)}_{(\mathbf{q}, r)}=\frac{\delta_{\mathbf{q},0}
}{N\hbar\beta}\sum_{\mathbf{k},n}{\rm
Tr}[\mathbf{G}_{\mathbf{k}}(i\hbar\omega_n)\mathbf{P}^{s}]e^{i\zeta
\omega_n }\eta_{sr}.\nonumber
\end{equation}

Recall that by definition $p_{0}=-\omega_n$. The form of the Green's
function allows us to perform the Matsubara summation, 
which yields
\begin{equation}
\lim_{\zeta\rightarrow 0+ }
\sum_{n}\mathbf{G}_{\mathbf{k}}(i\hbar\omega_n)e^{i\zeta \omega_n
}=\hbar\beta
\sum_{\alpha=1}^{4}\mathbf{U}_\mathbf{k}\mathbf{I}_{\alpha}
\mathbf{U}_\mathbf{k}^{\dag}n_F(\tilde{\varepsilon}_\mathbf{k}^{(\alpha)}),\nonumber
\end{equation}
where the Fermi distribution function $n_F(z)=(e^{\beta z }+1)^{-1}$
and the energy is measured with respect to the chemical potential
$\tilde{\varepsilon}_\mathbf{k}^{(\alpha)}=\varepsilon_\mathbf{k}^{(\alpha)}-\mu$.
Thus, the expression for  $\Gamma^{(1)}$ simplifies to
\begin{equation}\label{gamma_one}
\Gamma^{(1)}_{(\mathbf{q}, r)}=\frac{ \delta_{\mathbf{q},0}
}{N}\sum_{\mathbf{k},\alpha}T^{(1)\;
s}_{\mathbf{k},\alpha}n_F(\tilde{\varepsilon}_\mathbf{k}^{(\alpha)})\eta_{sr},\nonumber
\end{equation}
where
\begin{equation}\label{t_one}
T^{(1)\; s}_{\mathbf{k},\alpha}={\rm Tr}[\mathbf{U}_\mathbf{k}
\mathbf{I}_{\alpha} \mathbf{U}_\mathbf{k}^{\dag}\mathbf{P}^{ s}],\nonumber
\end{equation}
and the trace is now taken only over the $4\times 4$ matrices in the
spin space. Analogously, for $\Gamma^{(2)}$,
\begin{widetext}
\begin{equation}
\Gamma^{(2)}_{(\mathbf{q}, r),(\mathbf{q}', r')}=\frac{-1}{\hbar^2\beta N
}{\rm
Tr}[\mathbf{G}_{\mathbf{p}',\mathbf{p}+\mathbf{q}}\mathbf{P}^{s}
\mathbf{G}_{\mathbf{p},\mathbf{p}'+\mathbf{q}'}\mathbf{P}^{s'}]\eta_{sr}\eta_{s'r'}e^{-i\zeta
(p_{0}-q_{0}+p'_{0}-q'_{0})}.\nonumber
\end{equation}
For homogeneous systems, this simplifies to
\begin{equation}\label{gamma2_homog}
\Gamma^{(2)}_{(\mathbf{q}, r),(\mathbf{q}', r')}=-\frac{\delta_{\mathbf{q}+\mathbf{q}',0}}{\hbar^2\beta
N }\lim_{\zeta\rightarrow 0+ }\sum_{\mathbf{k},m}{\rm
Tr}[\mathbf{G}_{\mathbf{k}+\mathbf{q}}(i\hbar\omega_{m}+i\hbar\Omega_n)\mathbf{P}^{s}\mathbf{G}_{\mathbf{k}}(i\hbar\omega_{m})\mathbf{P}^{s'}]
e^{i\zeta \omega_m }\eta_{sr}\eta_{s'r'},
\end{equation}
with $\Omega_n=2\pi n/\hbar\beta$ the bosonic Matsubara frequency,
because it is the difference between two fermionic Matsubara
frequencies. Performing the Matsubara summation in Eq.\;(\ref{gamma2_homog}) yields
\begin{equation}
\Gamma^{(2)}_{(\mathbf{q}, r),(\mathbf{q}', r')}=-\frac{\delta_{\mathbf{q}+\mathbf{q}',0}}
{N}\sum_{\mathbf{k},\alpha,\beta}\left(\frac{n_F(\tilde{\varepsilon}_{\mathbf{k} +\mathbf{q}}^{(\alpha)})}
{\tilde{\varepsilon}_{\mathbf{k}+\mathbf{q}}^{(\alpha)}-\tilde{\varepsilon}_{\mathbf{k}}^{(\beta)}-i\hbar\Omega_n}
+\frac{n_F(\tilde{\varepsilon}_{\mathbf{k}}^{(\beta)})}
{\tilde{\varepsilon}_{\mathbf{k}}^{(\beta)}-\tilde{\varepsilon}_{\mathbf{k}+\mathbf{q}}^{(\alpha)}+i\hbar\Omega_n}\right)T^{(2)\;
s,s'}_{\mathbf{k}+\mathbf{q},\alpha;\mathbf{k},\beta}\eta_{sr}\eta_{s'r'},\nonumber
\end{equation}
where
\begin{equation}\label{fullfactor}
T^{(2)\;
s,s'}_{\mathbf{k}+\mathbf{q},\alpha;\mathbf{k},\beta}\equiv{\rm
Tr}[\mathbf{U}_{\mathbf{k}+\mathbf{q}} \mathbf{I}_{\alpha}
\mathbf{U}_{\mathbf{k}+\mathbf{q}}^{\dag}\mathbf{P}^{s}
\mathbf{U}_\mathbf{k}
\mathbf{I}_{\beta}\mathbf{U}_\mathbf{k}^{\dag}\mathbf{P}^{s'}],\nonumber
\end{equation}
or in the more conventional form
\begin{equation}
\Gamma^{(2)}_{(\mathbf{q}, r),(\mathbf{q}', r')}=-\frac{\delta_{\mathbf{q}+\mathbf{q}',0}}
{N}\sum_{\mathbf{k},\alpha,\beta}\frac{n_F(\tilde{\varepsilon}_{\mathbf{k}+\mathbf{q}}^{(\alpha)})-n_F(\tilde{\varepsilon}_{\mathbf{k}}^{(\beta)})}
{\tilde{\varepsilon}_{\mathbf{k}+\mathbf{q}}^{(\alpha)}-\tilde{\varepsilon}_{\mathbf{k}}^{(\beta)}-i\hbar\Omega_n}T^{(2)\;
s,s'}_{\mathbf{k}+\mathbf{q},\alpha;\mathbf{k},\beta}\eta_{sr}\eta_{s'r'}.\nonumber
\end{equation}
\end{widetext}

\end{document}